\documentclass{aa}
\usepackage[utf8]{inputenc}
\usepackage{amsmath}
\usepackage{graphicx}
\usepackage{siunitx}
\usepackage{comment}
\usepackage{xcolor}
\usepackage{multirow}
\usepackage[colorlinks=true, allcolors=blue]{hyperref}

\newcommand{\botrule}{\hline\hline}

\newcommand{\hu}{${\rm km/s/Mpc}\,$}
\newcommand{\pdetGW}{$p_{\rm det}^{\rm GW}\,$}
\newcommand{\pdetEM}{$p_{\rm det}^{\rm EM}\,$}
\renewcommand{\d}{{\rm d}}

\begin{document}

\title{The potential role of binary neutron star merger afterglows in multimessenger cosmology}

\author{S. Mastrogiovanni\thanks{mastrosi@apc.in2p3.fr}$^{1}$, R. Duque\thanks{duque@iap.fr}$^{2}$, E. Chassande-Mottin$^{1}$, F. Daigne$^{2}$, R. Mochkovitch$^{2}$}

\titlerunning{Multimessenger cosmology with afterglows}

\institute{$^1$ Université de Paris, CNRS, Astroparticule et Cosmologie, F-75006 Paris, 
France\\ $^2$ Sorbonne Université, CNRS, Institut d'Astrophysique de Paris, F-75014 Paris, France}

\authorrunning{Mastrogiovanni et al.}

\date{\today}

\abstract{
Binary neutron star mergers offer a new and independent means of measuring the Hubble constant $H_0$ by combining the gravitational-wave inferred source luminosity distance with its redshift obtained from electromagnetic follow-up. 
This method is limited by the intrinsic degeneracy between the system distance and orbital inclination in the gravitational-wave signal. 
Observing the afterglow counterpart to a merger can further constrain the inclination angle, allowing this degeneracy to be partially lifted and improving the measurement of $H_0$.
In the case of the binary neutron star merger GW170817, afterglow light-curve and imaging modeling thus allowed the $H_0$ measurement to be improved by a factor of three.
However, systematic access to afterglow data is far from guaranteed. In fact, though each one allows a leap in $H_0$ precision, these afterglow counterparts should prove rare in forthcoming multimessenger campaigns.
We combine models for emission and detection of gravitational-wave and electromagnetic radiation from binary neutron star mergers with realistic population models and estimates for afterglow inclination angle constraints. Using these models, we quantify how fast $H_0$ will be narrowed down by successive multimessenger events with and without the afterglow. We find that because of its rareness and though it greatly refines angle estimates, the afterglow counterpart should not significantly contribute to the measurement of $H_0$ in the long run.}

\maketitle

\section{Introduction}
\label{sec:1}
The detection of gravitational waves (GWs) from compact binary coalescence \citep{2019PhRvD.100j4036A} has opened a new window to study the Universe.
Gravitational-wave sources are a new type of ``standard candle'', usually referred to as ``standard sirens'' as it is possible to directly measure their luminosity distance ($d_L$) from the GW signal \citep{1986Natur.323..310S}.
Hence, if supplied with the source redshift information, GW detections can be used to measure cosmological parameters \citep{2005ApJ...629...15H,2013arXiv1307.2638N,2018Natur.562..545C,2019PhRvD.100j3523M}, such as the Hubble constant $H_0$. This possibility is of great interest given the current tension between the $H_0$ measurement at early and late epochs of the Universe \citep{2017NatAs...1E.169F}.

The first GW measurement of $H_0$ was made possible by the multimessenger observation of the binary neutron star (BNS) merger GW170817 \citep{2017ApJ...848L..12A} and its associated kilonova, which enabled the identification of the host galaxy and its redshift, leading to a new and independent measurement of $H_0=70^{+12}_{-8}$~\hu  \citep{2017Natur.551...85A}. Currently, GW170817 is the only GW event observed with an electromagnetic  counterpart that allowed this kind of cosmological measure. In the absence of an electromagnetic counterpart, one can use the redshifts of all cataloged galaxies with positions consistent with the 3D GW skymap to measure $H_0$ \citep[the so-called ``dark siren'' method;][]{2019ApJ...871L..13F,2019ApJ...876L...7S,2020PhRvD.101l2001G}, leverage tidal effects in the GW waveform to estimate the source redshift \citep{2012PhRvL.108i1101M,2017PhRvD..95d3502D}, or exploit the power spectrum of GWs and galaxy distributions \citep{Mukherjee:2020hyn,Mukherjee:2019wcg}.

These methods encounter challenges due to galaxy catalog incompleteness and the difficulty in detecting tidal effects in the GW signal, as well as uncertainty in the equation of state of neutron star matter. Overall, an electromagnetic counterpart detection remains the best prospect for obtaining precise $H_0$ measurements with GWs.

The $H_0$ measurement is limited by the large uncertainty on the GW luminosity distance. This arises because the GW emission is not isotropic and the luminosity distance determination is degenerate with the binary orbital inclination to the line-of-sight, hereafter referred to as $\iota$ \citep{2019PhRvX...9c1028C}. To improve the $H_0$ measurement, it is crucial to break this degeneracy.

One solution is to measure $\iota$ from anisotropic electromagnetic signals emitted after the merger.
The most stringent constraint comes from the photometry and very long baseline interferometry (VLBI) imaging of the afterglow. This is nonthermal radiation from the shock formed as the relativistic merger outflow decelerates in the circum-merger environment. For GW170817, these observations provided inclination angle measurements precise to $\sim 12$~deg \citep[e.g.,][]{2017ApJ...851L..36G, TVRRB+2019, HMAKB+2019} and $\sim 5$~deg, respectively \citep{MDGNH+2018,CDMM2019,GSPGY+2019,2020arXiv201104001A}, and refined the $H_0$ measurement to $H_0=70.3^{+5.3}_{-5.0}$ \hu~\citep{2019NatAs...3..940H}.

Further afterglow counterparts could drastically improve $H_0$ measurements and play a leading role in multimessenger cosmology.
However, they are faint and difficult to detect for distant or very inclined binaries and should therefore prove rare in the future \citep{2018MNRAS.475..699S,GNP2019,DDM2019}. Furthermore, using electromagnetic measurements of $\iota$ entails selection effects that, if not correctly understood, can significantly bias the subsequent measurement of $H_0$ and must be carefully taken into account \citep{2020arXiv200602779C}.

In this prospective study, we consider both the likeliness of the afterglow to be observed and its capability to constrain $\iota$ to realistically quantify its benefit for the measurement of the Hubble constant with future multimessenger events. We will assume that the peculiar motion of the galaxies is accurately known, as their mismatch could introduce a systematic bias.

Afterglow-derived information on $\iota$ is much more precise than that from GW data only. Nonetheless, we find that, for advanced GW detectors in their O3 and O4+ configurations, detecting afterglows should be so rare that, statistically, afterglows will not accelerate the narrowing-down of $H_0$.

This paper is organized as follows.
In Sect.~\ref{sec:2} we describe the possible multimessenger observation scenarios of a GW event with electromagnetic counterparts following which a measurement of $H_0$ can be made. In Sect.~\ref{sec:3} we evaluate and quantify the likeliness of the different scenarios by combining models for the emission of GWs and electromagnetic radiation from BNSs, models for the detection of these signals, and population models for the sources. In Sect.~\ref{sec:4}, we briefly describe the prospects of joint GW and electromagnetic detections in these different scenarios. In Sect.~\ref{sec:5}, we evaluate the potential of electromagnetic counterparts in contributing to the measurement of $H_0$ by combining their capability to measure the inclination angle $\iota$ with their likelihood of being observed in the scenarios described above.

\section{Inferring $H_0$ with gravitational waves and electromagnetic counterparts}
\label{sec:2}

In this section, we review the possible multimessenger observation scenarios expected after a GW event from a BNS merger.

Until the advent of deep radio surveys such as the Square Kilometer Array \citep{DHSL2009}, the kilonova will continue to be the only electromagnetic counterpart to GW events that can lead to an identification of the system host galaxy and redshift. For a kilonova to be detected during follow-up of a GW event, it must be in reach of the follow-up telescopes in terms of both magnitude and sky position, and the source must be localized sufficiently well to be discovered in time before it fades. Magnitude and localization are, however, not the only conditions for detection. 
As recent searches for kilonova signals both during GW follow-up campaigns \citep{KAASS+2020} and in archival data \citep{AKSKB+2020} have shown, difficulties can arise in recognizing kilonovae among a myriad of optical transients, even with quality spectroscopic or color evolution observations.

Detection of the kilonova counterpart and thereby acquiring the system's redshift through its host galaxy is the minimal scenario required for a multimessenger measurement of $H_0$. We refer to this scenario as ``Level 1.'' It is subject to detection criteria concerning the GW and kilonova signals, described in Sect.~\ref{sec:3}.

In this scenario, the information on $d_L$ is provided solely by the GW data, without any contribution from electromagnetic counterparts. Indeed, any direct distance information from an electromagnetic counterpart would require using the cosmic distance ladder, which is of course excluded in the perspective of measuring $H_0$.

In principle, the kilonova signal could indirectly contribute through the measurement of $\iota$ from color-evolution considerations \citep{KRA2019,DBGSS2020}. However, currently, these signals crucially lack modeling and observing history. 
Inclination angle measurements from kilonova data are very model dependent and lack robustness \citep{2020ApJ...892L..16D,2021MNRAS.502.3057H}. Only when the kilonova sample grows will the potential impact of kilonova-derived angle constraints be appreciated. Therefore, we excluded any contribution to $\iota$ from kilonova data.

As expected from the observation of short gamma-ray bursts and evidenced in the case of GW170817, relativistic jets are launched from BNS mergers \citep{MDGNH+2018, GSPGY+2019}. The interaction of this jet with the circum-merger environment leads to long-lived\footnote{The afterglow of GW170817 was still detected in the X-ray band 1000 days post-merger \citep{HMLAF+2020}.} nonthermal emission in the radio to the X-ray bands: the afterglow counterpart. 

The afterglow photometry can provide an independent measurement of inclination angle. In such measurements from off-axis jet afterglow data fitting, there is unavoidable degeneracy with the jet opening angle \citep{NP2020}. However, when combined with prior estimates for short gamma-ray burst jet opening angles (see references in Sect.~\ref{sec:3.2.3}), afterglow data can still lead to inclination angle information. It can therefore indirectly inform on $d_L$ by breaking the $d_L/\iota$ degeneracy in the GW data. 
We refer to the scenario where Level 1 is realized and a $\iota$ measurement from the jet afterglow photometry is made as ``Level 2.''
It is subject to the realization of the GW and kilonova detection criteria and to detection criteria on the jet afterglow light curve, which we detail in the next section.

Additionally, the relativistic nature of the jetted outflow can be revealed by VLBI observations \citep{MDGNH+2018, GSPGY+2019} that evidence an apparent superluminal motion of the jet head.
Detecting this centroid displacement is possible for events that are particularly close or bright or under specific inclination angle conditions \citep{DDM2019,DKHMD+2020}. Doing so further constrains $\iota$ and narrows down the measurement of $H_0$, as shown in the case of GW170817. 
We refer to the scenario where such a constraint on $\iota$ can be extracted from afterglow VLBI imaging, in addition to those of afterglow photometry, as ``Level 3.'' This final level of $\iota$ constraint is the most informative on $\iota$ and $H_0$, but also the most difficult to obtain.

Other electromagnetic counterparts could potentially provide further independent measurements of $\iota$. These are the short gamma-ray burst and the rebrightening in the source's multiwavelength signal due to the emergence of emission from the front shock of the decelerating mildly relativistic ejecta responsible for the kilonova emission, called the ``kilonova afterglow'' \citep{HKSNP2018, KGB2019}. 
However, the short gamma-ray burst should prove extremely rare in future events \citep{Ghirlanda:2016ijf,BPBG2019} and there is a lack of robust modeling for gamma-ray signals, especially for very inclined events. Furthermore, the singularity of GRB170817A with respect to other short gamma-ray bursts has cast more uncertainty on the modeling of gamma-ray emission from BNS mergers \citep{KNSKC+2017,NGPKH2018}.
Modeling of the kilonova afterglow is still uncertain, rendering any angle measurement difficult \citep{DBDM2020}. To sum up, we consider neither the gamma-ray burst nor the kilonova afterglow as viable to measure $\iota$ and thus $H_0$.

\section{Evaluation of detection probabilities and selection effects}
\label{sec:3}

Here, we evaluate the likelihoods of the different observing scenarios to occur, by computing the probabilities of detection of the GW and electromagnetic signals.

\subsection{Gravitational-wave detection probability}
\label{sec:3A}
In GW searches an event is detected if its detected signal-to-noise ratio (S/N) $\hat\rho_{\rm det}$ exceeds a certain threshold. The detected S/N is a measure of the power of the GW signal registered at the GW detectors. The higher the S/N, the more chance there is to recognize the signal from detector noise.
It is important to note that $\hat\rho_{\rm det}$ differs from the optimal-filter S/N $\rho_{\rm opt}$ that is calculated taking into account the average sensitivity of the detector network. 
Indeed, noise fluctuations are not included in optimal-filter S/N. 
Also, due to the same noise fluctuations, the GW event will be detected with values $\hat d_L$ and $\cos \hat \iota$ that differ from the true $d_L$ and $\cos \iota$.
The detected $\hat d_L, \cos \hat \iota$, and the corresponding $\hat \rho_{\rm det}$ can be sampled from the GW likelihood model.

In this article, we used the Cutler and Flanagan (CF) approximant for the GW likelihood \citep{1994PhRvD..49.2658C,1995PhRvD..52..848P,2019PhRvD.100h3514C}. 
This approximant can reproduce the GW likelihood in the $\cos \hat \iota$ and $\hat d_L$ space under the assumption that the chirp mass of the signal is well estimated, as is always the case for BNS detections \citep{1994PhRvD..49.2658C}.
By sampling from the CF approximate, we can obtain a value of the detected $\cos \hat \iota$ and $\hat{d_L}$, which can then be used to compute the detected S/N $\hat{\rho}_{\rm det}$ with the following \citep{1994PhRvD..49.2658C,2019PhRvD.100h3514C}:
\begin{equation}
    \hat{\rho}^2_{\rm \det}  \equiv  \rho_{\rm fo}^2  {\sigma_d} \Big[  (\chi_+^2 + \cos^2 \hat{\iota})+  \epsilon_d  (\chi_+^2 - \cos^2 \hat{\iota} )\cos(4\bar{\psi})\Big],
    \label{eq:SNR}
\end{equation}
where $\chi_+ \equiv (1+\cos^2 \hat{\iota})/2$, while $\epsilon_d, \sigma_d$ are variables that depend on the detector network and the sky-position of the GW source and $\bar{\psi}$ is the GW polarization angle. The variable $\rho_{\rm fo}$ is the optimal-filter S/N that the binary would have had if it had been face-on,
\begin{equation}
    \rho_{\rm fo}^2 = 
    \frac{1}{\hat d_L^{2}}\left[ \frac{5}{6\pi^{4/3}}\frac{G{\mathcal M}_{c}^{5/3}}{c^3}\int_{f_{\rm low}}^{f_{\rm LSO}} \mathrm{d}f \dfrac{f^{-7/3}}{S_{n,aver}(f)} \right]\:,
    \label{eq:SNR_fo}
\end{equation}
with $S_{n,aver}(f)$ the harmonic mean of the noise power spectrum densities (PSD) of the interferometric detectors composing the network, $G$ and $c$ universal gravitational and speed-of-light constants and $\mathcal{M}_c$ the system's chirp mass in the detector frame. As per the low redshift range of the systems in this study (see below), we assimilate the chirp mass and last stable orbit frequency $f_{\rm LSO}$ in the detector frame to their values in the source frame. The integral low boundary $f_{\rm low} = 20$~Hz is set to the low-frequency cut-off for ground-based GW detectors.

To evaluate the GW detection probability, we perform a Monte Carlo simulation with synthetic signals. 
We divide the $\cos \iota$ range into 200 bins and we simulate 10000 BNS merging at fixed $d_L$ and uniformly distributed over the celestial sphere.  
The BNS masses are generated from a Gaussian distribution with a mean of $1.35 M_\odot$ and a standard deviation of $0.15 M_\odot$ \citep{2019ApJ...876...18F}.
For each binary, we then draw a detected $\hat d_L$ and $\cos \hat \iota$ from the CF likelihood and calculate $\hat \rho_{\rm det}$ following Eqs.~\ref{eq:SNR}-\ref{eq:SNR_fo}. We then count the binaries with S/N exceeding the online match-filtering threshold for detection of $14$\footnote{Usually a S/N threshold of 8 is assumed for the detection by a single detector \cite{2018LRR....21....3A}. As the CF approximant is valid in the high S/N regime, here we assume a threshold of 14. This is equivalent, on average, to a S/N of about 8 in each detector.} and compute the detection probability as the fraction of BNS events detected.
The previous procedure is repeated in $d_L$ with a step of $1$~Mpc until only 1 of the 10000 simulated systems is detected. 
Following \cite{2017arXiv170908079C}, we define the 0.02\% ``response distance'' $d^{r}_{0.2\%}$ at which 0.2\% of the simulated binaries (with isotropic distribution in the sky and orientation) will be detected by the network.
We verified by running the simulation several times that the GW detection probability estimation is not prone to Monte Carlo statistical fluctuations.

For this paper, we use three PSDs for the HLV network, composed of the LIGO Hanford, LIGO Livingston, and Virgo detectors. The first PSD is indicative of the detectors' sensitivity during O2 \citep{2017PhRvL.119p1101A}, the second is indicative of the sensitivity reached during the first 3 months of O3 \citep{2018LRR....21....3A}, while the third one is a projection for the O4 run sensitivity \citep{2019PASP..131b4503B}.

\begin{figure}
    \centering
    \includegraphics[width=\linewidth]{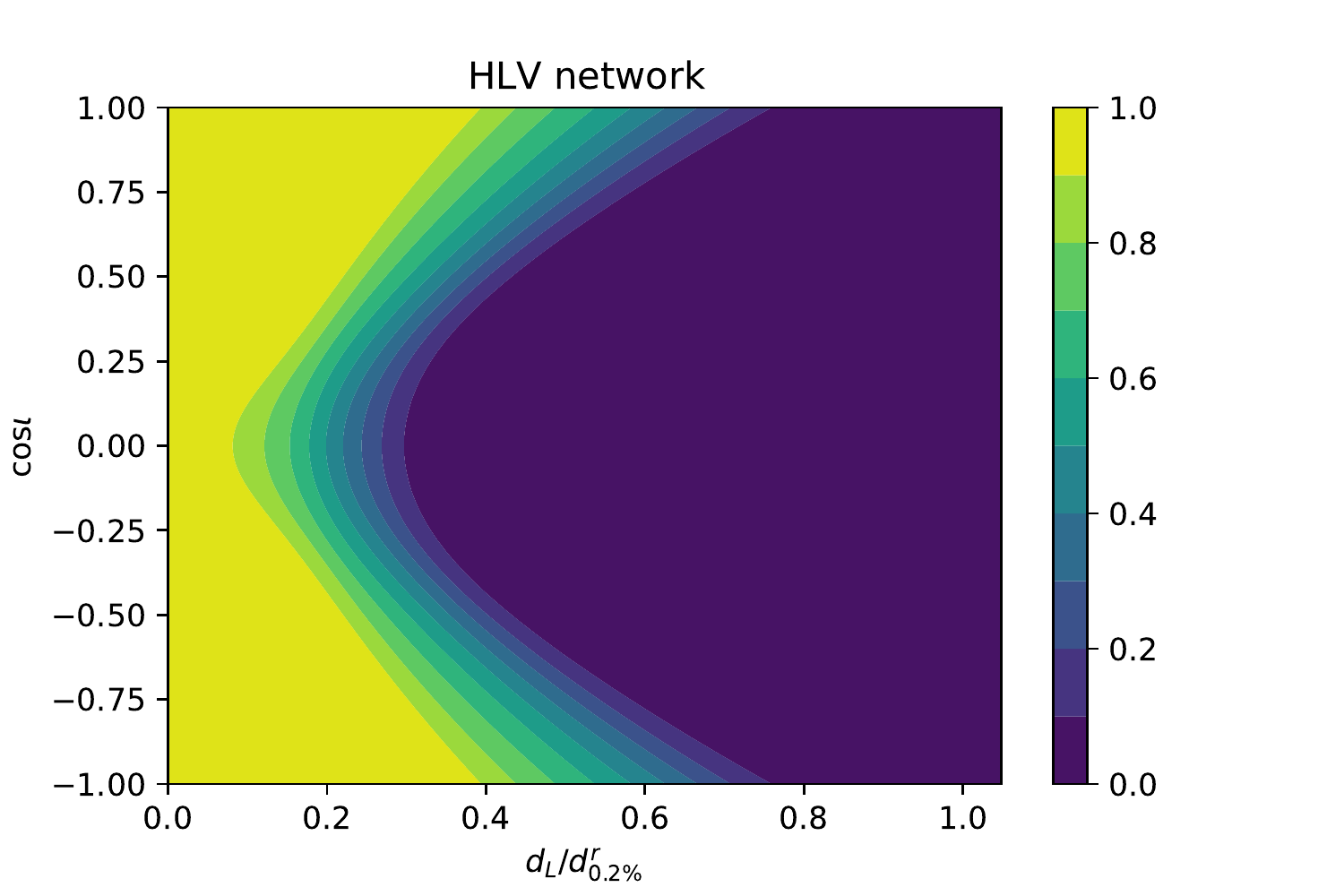}
    \caption{GW detection probability \pdetGW as a function of the BNS luminosity distance and inclination for the detector network HLV using a S/N threshold of 14.
    The horizontal axis is scaled to the BNS 0.2\% response distance $d^{r}_{0.2\%}$, which is $d^{r}_{0.2\%}=107$ Mpc for O2; $d^{r}_{0.2\%}=218$ Mpc for O3; $d^{r}_{0.2\%}=477$ Mpc for O4.
    }
    \label{fig:pdetGW}
\end{figure}

Fig.~\ref{fig:pdetGW} shows the GW detection probability as a function of the BNS $d_L$ and $\cos \iota$ marginalized over the GW polarization angle and sky-position. The systems we consider have redshifts $\leq 0.06$ for both Planck \citep{PAAAA+2018} and SH0ES \citep{RCYMS2019} values of $H_0$. Because of the low redshift range of our simulation, the shape of the GW detection probability functions are not affected by the nonlinear $d_L$-$z$ relationship and are therefore the same for all the PSD hypotheses. As it can be seen from the figure, for all three GW sensitivity levels, face-on binaries are easier to detect as the GW emission is stronger perpendicularly to the orbital plane, and they can be observed at higher luminosity distances. 

\subsection{Electromagnetic counterpart detection probability}
\label{sec:3B}

\begin{table}[t!]
\caption{Description and numerical values of various constants used in the electromagnetic emission and detection models. }
\begin{tabular}{|>{\centering}p{29mm}|p{50mm}|}
\hline
Notation & Description \\ \hline 
$G_0 = -17.6$ & Peak absolute magnitude of a face-on kilonova in the $g$ band \\ \hline
$R_0 = -16.9$ & Same, in the $r$ band \\
\hline
$\theta_v^{\rm max, KN} = 60\deg$ & Viewing angle after which the kilonova magnitude varies no more with viewing angle, in both $r$ and $g$ bands\\
\hline
$\Delta G = 7$ & magnitude contrast between a face-on and a $\theta_v^{\rm max, KN}$-angle view of a kilonova in the $g$ band \\ \hline
$\Delta R = 4$ & Same, in the $r$ band \\ \hline
$p_{\rm night} = 0.52$ & Fraction of the entire sky accessible to a $\sim 33\deg$-latitude follow-up instrument in a single night \\ 
\hline
$r_{\rm lim}, g_{\rm lim} = 21$ & Limiting magnitude of the optical follow-up instruments \\
\hline
$F_{\rm lim} = 15~\mu$Jy & Flux sensitivity of the radio follow-up instruments \\ 
\hline
$\theta_j = 0.1$ rad & Half-opening angle of the jet launched by the merger \\
\hline
$\Delta \theta_{{\rm VLBI}} = 2$ mas & Angular resolution of the VLBI array \\ \hline \botrule
\end{tabular}
\label{tab:numbers}
\end{table}

\subsubsection{Level 1 observing scenario}

In the perspective of upcoming high-cadence and large-field-of-view optical facilities such as the Zwicky Transient Facility \citep[ZTF,][]{BKGDS+2019} or Vera C. Rubin Observatory \citep[LSST,][]{IABBC+2008}, we assume the detection of the kilonova is limited only by the kilonova magnitude and sky-position, and not by the size of the GW-provided skymap that the follow-up network must cover in its searches.
Indeed, we consider that these survey facilities can cover all the sky available to them within the first nights of the search, that is, before the estimated time for significant dimming of kilonova signals.  

While this can be considered a ``best-case'' assumption, this level of performance was reached during the campaigns following GW events in the recent O3 run of the LIGO-Virgo Collaboration by, for example, the GROWTH collaboration \citep{KAASS+2020}. In particular, for the only confirmed BNS event of the O3 run GW190425 \citep{CAADH+2019}, the ZTF covered the $\sim 8000 \deg^2$ of the skymap overlapping with their night sky.

Similarly to the evaluation of GW detection probability, we place systems uniformly in the sky and uniformly in $\cos \iota$ and $d_L$ out to $600$ Mpc. 
This maximum distance ensures that a fraction of less than $10^{-4}$ of kilonovae were detected in the furthest distance bin. Our model for the inclination-angle-dependent peak magnitude of the kilonova signal is an empirical fit to model ``W2'' in \cite{WKFRE+2018} (their Fig.~19). It appears that the peak magnitudes are approximately linear in $\cos \iota$ up to a maximum angle $\theta_v^{\rm max, KN} \sim 60\deg$ after which the dependence on angle strongly decreases. We therefore set the peak absolute magnitude of these events as:
\begin{equation}
\left\{\begin{array}{ll}
G & = G_0 + \Delta G \frac{1 - \cos {\rm min}(\iota, \theta_v^{\rm max, KN})}{1 - \cos \theta_v^{\rm max, KN}}\\
R & = R_0 + \Delta R \frac{1 - \cos {\rm min}(\iota, \theta_v^{\rm max, KN})}{1 - \cos \theta_v^{\rm max, KN}}\end{array}\right.
\end{equation}
with numerical values reported in Table~\ref{tab:numbers}. We note that a similar dependence of kilonova magnitudes on $\iota$ was considered in \citet{VGBMC+2017} (their ``asymmetric model'') and found in \citet{2020ApJ...889..171K}. We select the detectable events as those satisfying the magnitude threshold criterion $g < g_{\rm lim}$ or $r < r_{\rm lim}$, with lowercase letters denoting the apparent magnitudes.

Among all the events with large enough flux, follow-up can only detect those in its accessible sky. For observatories at latitudes of $\sim 33\deg$ such as the ZTF (northern hemisphere) and the LSST (southern hemisphere), this represents a season-averaged fraction $p_{\rm night} \sim 52\%$ of the whole sky \citep{B2016}. 
Thus, we deem detected the remaining events with a coin-toss with probability $p_{\rm night} = 0.52$.

Forming the ratio with the original number of events allows us to evaluate the detection probability $p_{{\rm det}}^z$, the probability of acquiring an event redshift as a function of its luminosity distance and inclination angle. This is plotted in Fig.~\ref{fig:pdetEM}. 
Here, the detection probabilities were normalized to $p_{\rm night}$. As the $p_{\rm det}$ are even in $\iota$, we represent only the $\cos \iota > 0$ range. Following what was developed in Sect.~\ref{sec:2} and given the independence of the GW and kilonova detection processes, we define the probability of the Level 1 scenario occurring as $p^{{\rm L}_1} = p_{{\rm det}}^{\rm GW} \times p_{{\rm det}}^z$.

\subsubsection{Level 2 observing scenario}
\label{sec:3.2.2}
For events where emission from the core jet dominates the afterglow radiation, the afterglow light curves are expected to display a single peak occurring when the jet has decelerated to a Lorentz factor of $\Gamma \sim 1/\iota$ (e.g., \citealt{GG2018, MFDLC+2018}, but see \citealt{BGG2020} for a study of multipeaked afterglows). This peak can occur up to hundreds of days after the merger \citep{DDM2019}. Assuming instrument availability and long-term follow-up efforts, detecting the afterglow is simply a matter of flux sensitivity, once the source's position is settled by the kilonova.
However, detecting the afterglow at its peak does not suffice to make an inclination angle measurement. This requires an extended and well-sampled light curve on which to fit afterglow models, as was extensively done for GW170817 \citep[e.g.,][]{2017MNRAS.472.4953L,RSIMB+2018,PhysRevLett.120.241103,TVRRB+2019}. 
We therefore define the criterion for $\iota$ measurement with the afterglow light curve as:
\begin{equation}
F_p > 10 \times F_{\rm lim},
\label{eq:ag}
\end{equation}
where $F_p$ is the peak flux of the afterglow light curve, and $F_{\rm lim}$ is the limiting sensitivity of the follow-up facility. In this study, we consider the 3~GHz band and the Very Large Array (VLA) as the limiting radio facility, with $F_{\rm lim} = 15~\mu$Jy.

In addition to $\iota$ and $d_L$, the peak flux $F_p$ of every event depends on the jet's kinetic energy, the particle density in the surrounding medium, and on the microphysical parameters of the front shock formed by the decelerating jet, such as the spectral index of the shock-accelerated population of electrons, denoted $p$. The analytical form for $F_p$ as a function of these parameters that we use can be found in \cite{NPG2002}. 
There is some uncertainty in the distributions of these parameters in the population of jets from BNS mergers, in particular for the jet kinetic energy. 
To establish $p_{{\rm det}}^{z+{\rm AG}}$, the probability of making the inclination measurement allowed by the afterglow photometric data in addition to acquiring the redshift, we use the same distribution of parameters as the population model of \cite{DDM2019}. 
In particular, we use two distinct hypotheses regarding the distribution of the jets' kinetic energies. 
These are labeled G16 and WP15 in the sequel and are derived by assuming a constant conversion factor between the luminosity of a short gamma-ray burst and the post-burst jet kinetic energy, and starting from the short gamma-ray bursts luminosity functions found by \cite{GSPGS+2016} and \cite{WP2015}, respectively. Among published short gamma-ray burst luminosity functions, these two represent extremes in the steepness of the luminosity function, with G16 predicting many more bright bursts (and therefore many more energetic afterglows) than WP15.

By applying this population model to those events selected to establish $p_{{\rm det}}^z$ and applying the detection criterion in Eq.~\ref{eq:ag}, we calculate $p_{{\rm det}}^{z+{\rm AG}}$. It is plotted in Fig.~\ref{fig:pdetEM}. We define the likelihood that a Level 2 scenarios occurs as $p^{{\rm L}_2} = p_{{\rm det}}^{\rm GW} \times p_{{\rm det}}^{z+{\rm AG}}$. 
As expected, the range of parameter space allowed by this scenario is much smaller than for a kilonova-only event. In particular, the distant or inclined events are largely cut off because $F_p \propto \iota^{-4.4}/d_L^2$ \citep[for a shock-accelerated population spectral index of $p = 2.2$,][]{NPG2002}, whereas $r, g \sim 1 - \cos \iota + \log d_L$.

We note that GW170817 does not exactly qualify for our Level 2 scenario. Indeed, it had $\log F_p/F_{\rm lim} \sim 0.9 < 1$ \citep{MFDLC+2018}. 
It seems that much of the uncertainty in the measurement of $\iota$ with GW170817 is held in the very early phases of the afterglow, where the fitting models most diverge \citep{GSPGY+2019}. Had the afterglow been brighter--at the level of our Level 2 scenario--and these earlier points observed, a better measurement of $\iota$ would have certainly ensued. Nonetheless, as we detail in Sect.~\ref{sec:5}, we consider the case of GW170817 as representative of the $\iota$ measurements possible in the Level 2 scenario.

\subsubsection{Level 3 observing scenario}
\label{sec:3.2.3}
The Level 3 scenario occurs if the motion of the jet can be measured by VLBI imaging, in addition to the measurements of afterglow photometry. Once again this assumes constant instrument availability and long-term follow-up. We estimate the angular displacement of the jet centroid as:
\begin{equation}
\delta \theta_{r} = \delta t_{{\rm VLBI}} \times \frac{{\rm d}\theta}{{\rm d}t}_{|{\rm max}},
\end{equation}
where $\delta t_{{\rm VLBI}}$ is the total time the afterglow remains detectable by the radiotelescope array, that is, the time its flux is above $F_{\rm lim}$\footnote{To simplify, we assumed the radio flux threshold for acquiring the afterglow light curve and the jet motion imaging were the same, at the level of the VLA.}, and where ${\rm d}\theta/{\rm d}t|_{\rm max}$ is the proper motion of the remnant at the time of the afterglow peak, that is when $\Gamma \times \iota \sim 1$. At this time, it is straightforward to estimate the remnant's proper motion as:
\begin{equation}
\label{eq:ppm}
\frac{{\rm d}\theta}{{\rm d}t}_{|{\rm max}} \sim \frac{c\beta_{\rm app}}{d_L}
\end{equation}
with $\beta_{\rm app} = \frac{\beta \sin \iota}{1 - \beta \cos\iota}$ the apparent velocity of the remnant, $\beta = \sqrt{1 - 1/\Gamma_p^2}$ and $\Gamma_p = 1 / \iota$ the jet head Lorentz factor at the afterglow peak. By considering the source's proper motion to be that at afterglow peak during the entire follow-up, Eq.~\ref{eq:ppm} in fact over-estimates $\delta \theta_r$. Our Level 3 events are therefore likely treated in an optimistic manner, as we discuss in Sect.~\ref{sec:6}.

Also, Eq.~\ref{eq:ppm} is not valid if the observer is within the jet's opening, when $\iota < \theta_j$ with $\theta_j$ the half-opening angle of the jet. In this case, no jet displacement is observed, $\delta \theta_r = 0$. We consider $\theta_j = 0.1 \sim 6 \deg$, in line with measurements on GW170817 \citep{GG2018,VRRBW+2018,MFDLC+2018} and other short gamma-ray burst studies \citep{FBMZ2015,BPBG2019}.

The afterglow flux will remain in reach of the radio network for a duration that depends on the details of its light curve and therefore on, for example, the jet structure, its expansion dynamics, and the surrounding medium density profile. 
To simplify, we assume all jets launched from mergers have the same structure as GW170817 and all mergers occur in a rarefied medium with a constant-density profile as suggested by short gamma-ray burst observations. 
In this case, the slopes of the increasing and decreasing phases of the light curve are the same as for GW170817's afterglow, regardless of inclination angle \citep{BGG2020}. Therefore, we empirically modeled the afterglow light curves as a broken power-law with slopes $+0.80$ and $-2.2$ \citep{MFDLC+2018} respectively before and after a peak occurring at a time $T_p$ and a flux $F_p$. The time $T_p$ depends on the same parameters as $F_p$, and relevant equations can be found in \cite{NPG2002}. The time the signal is above the radio threshold can thus be analytically estimated, and thereby the total source displacement. Then, detection of this displacement is simply conditioned by the VLBI array angular resolution: $\delta \theta_{r} > \Delta \theta_{{\rm VLBI}} \sim 2~{\rm mas}$ \citep{GSPGY+2019}.

\begin{figure*}
    \centering
    \begin{tabular}{cccc}
     &\multicolumn{2}{c}{\hspace{4.2cm} \includegraphics[width=0.6\linewidth]{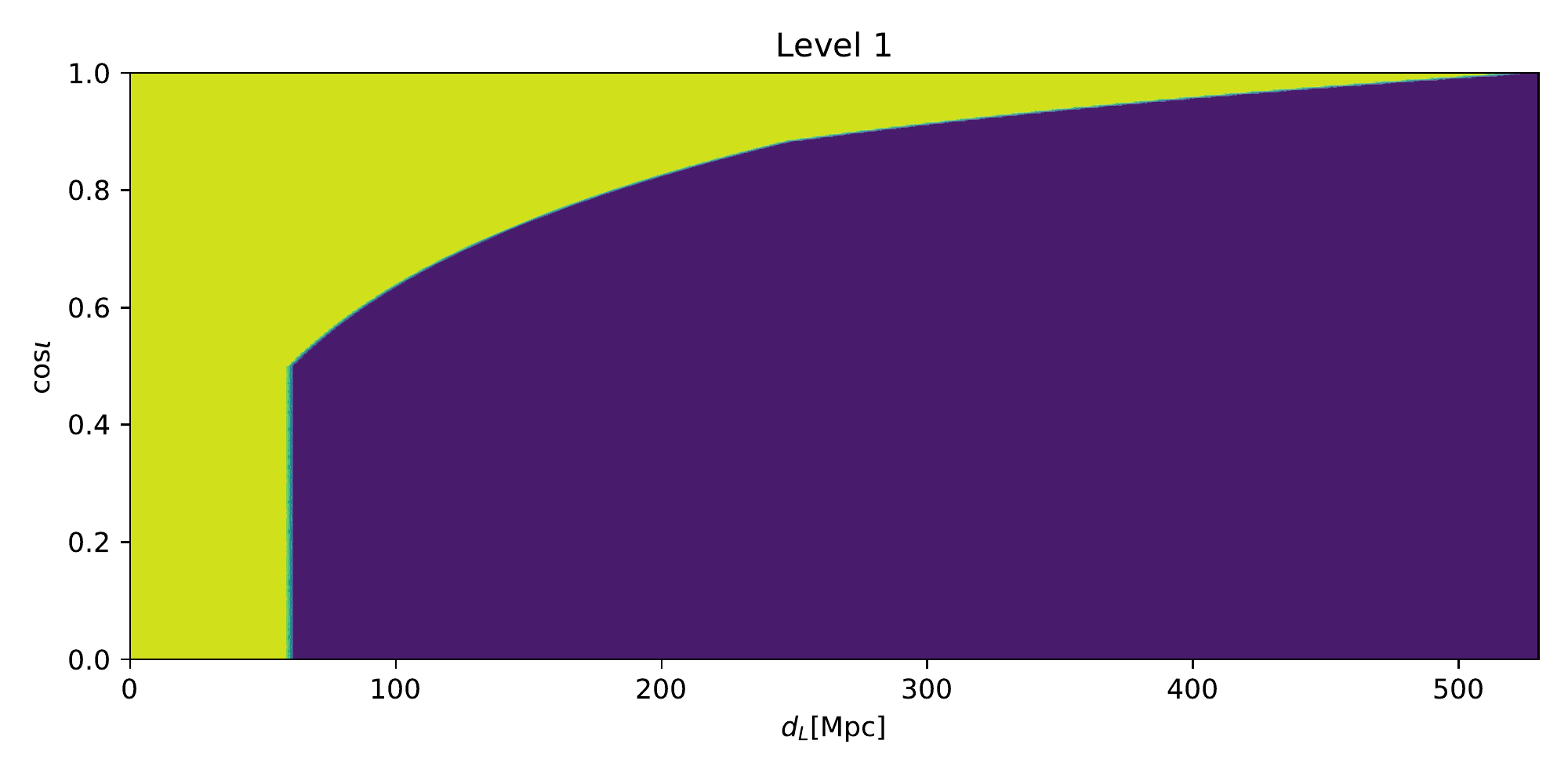}}& \\ \multicolumn{2}{c}{\includegraphics[width=0.48\linewidth]{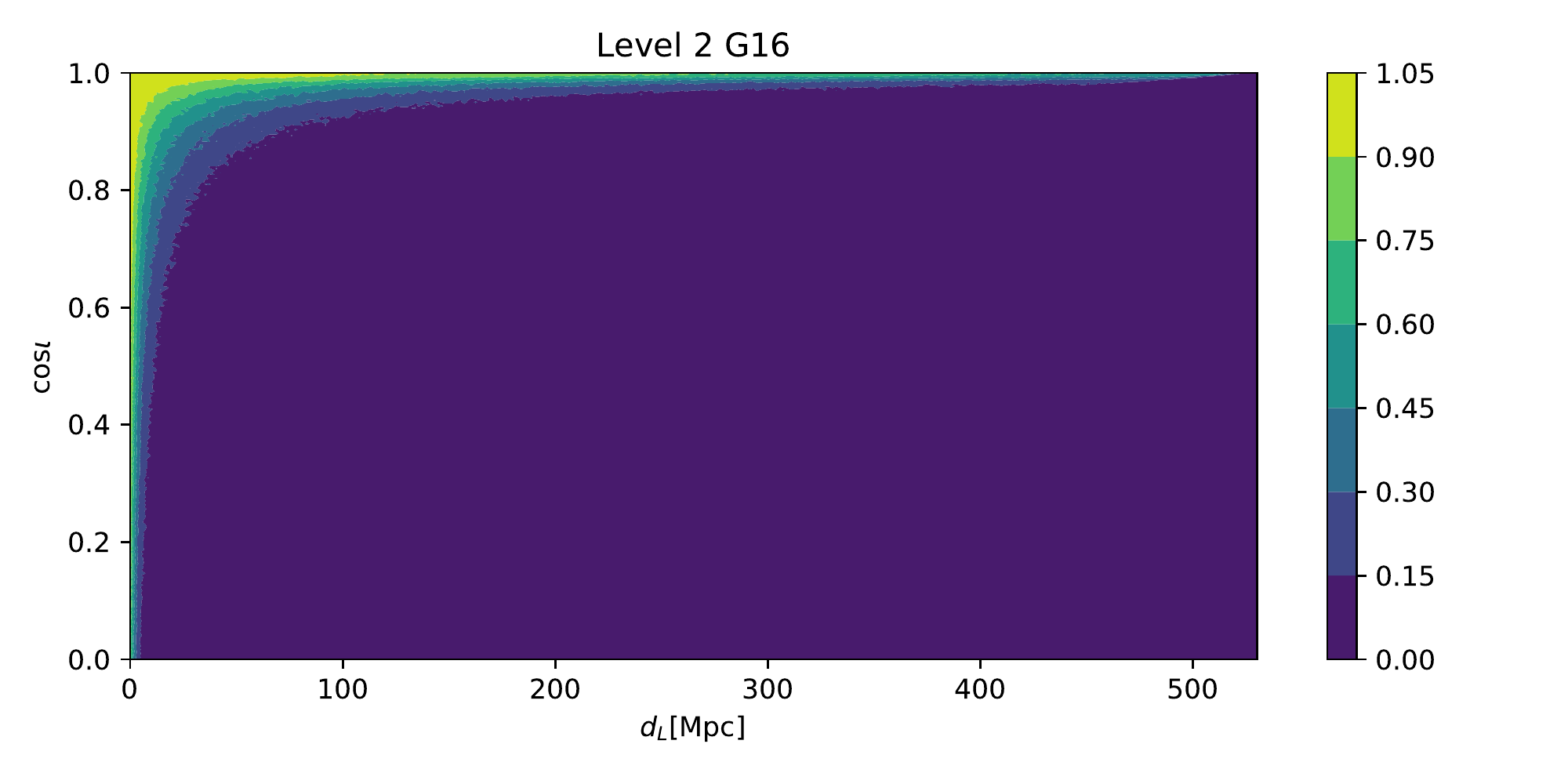}} & \multicolumn{2}{c}{\includegraphics[width=0.52\linewidth]{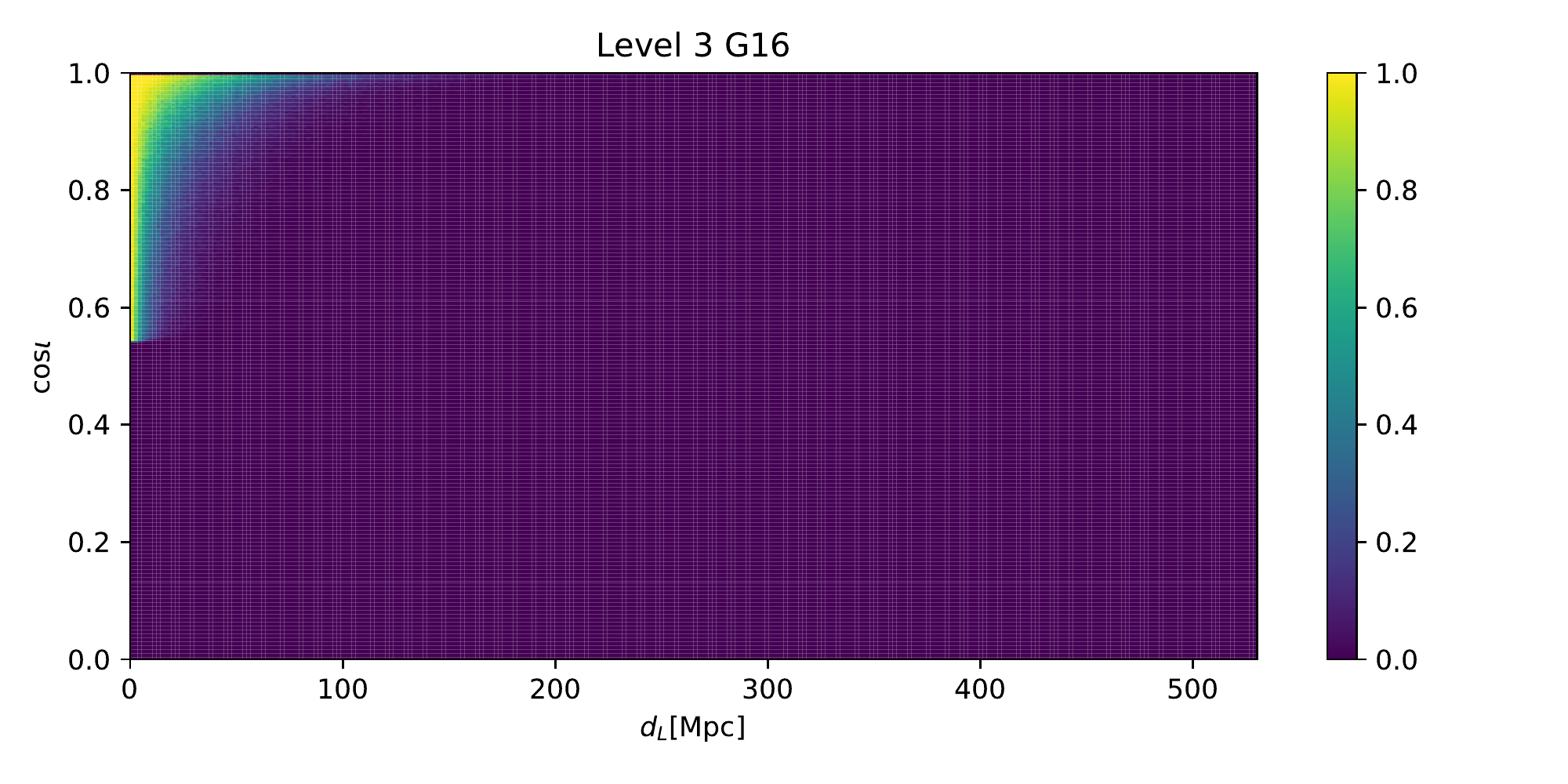}} \\
      \multicolumn{2}{c}{\includegraphics[width=0.48\linewidth]{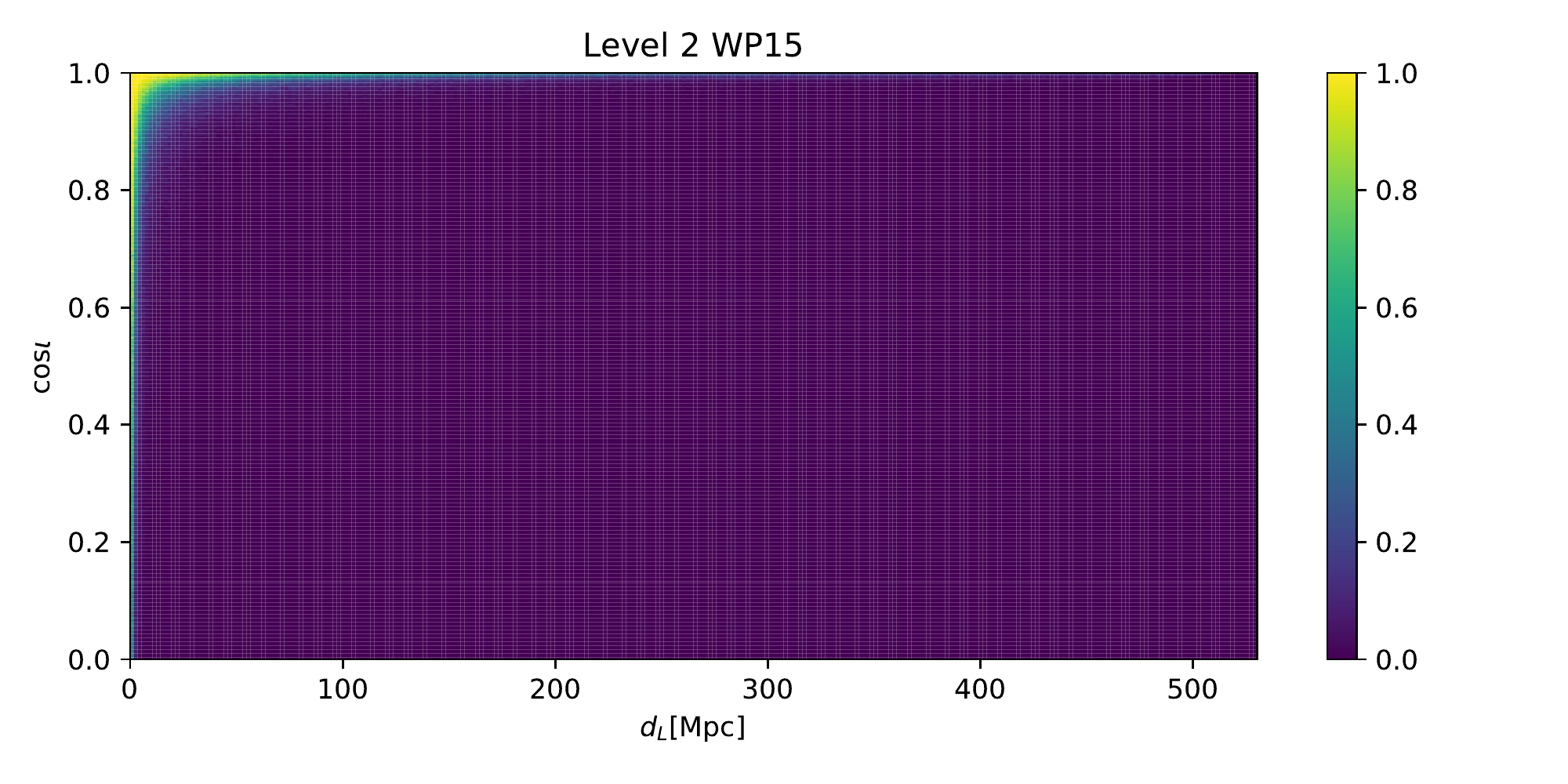}} & \multicolumn{2}{c}{\includegraphics[width=0.52\linewidth]{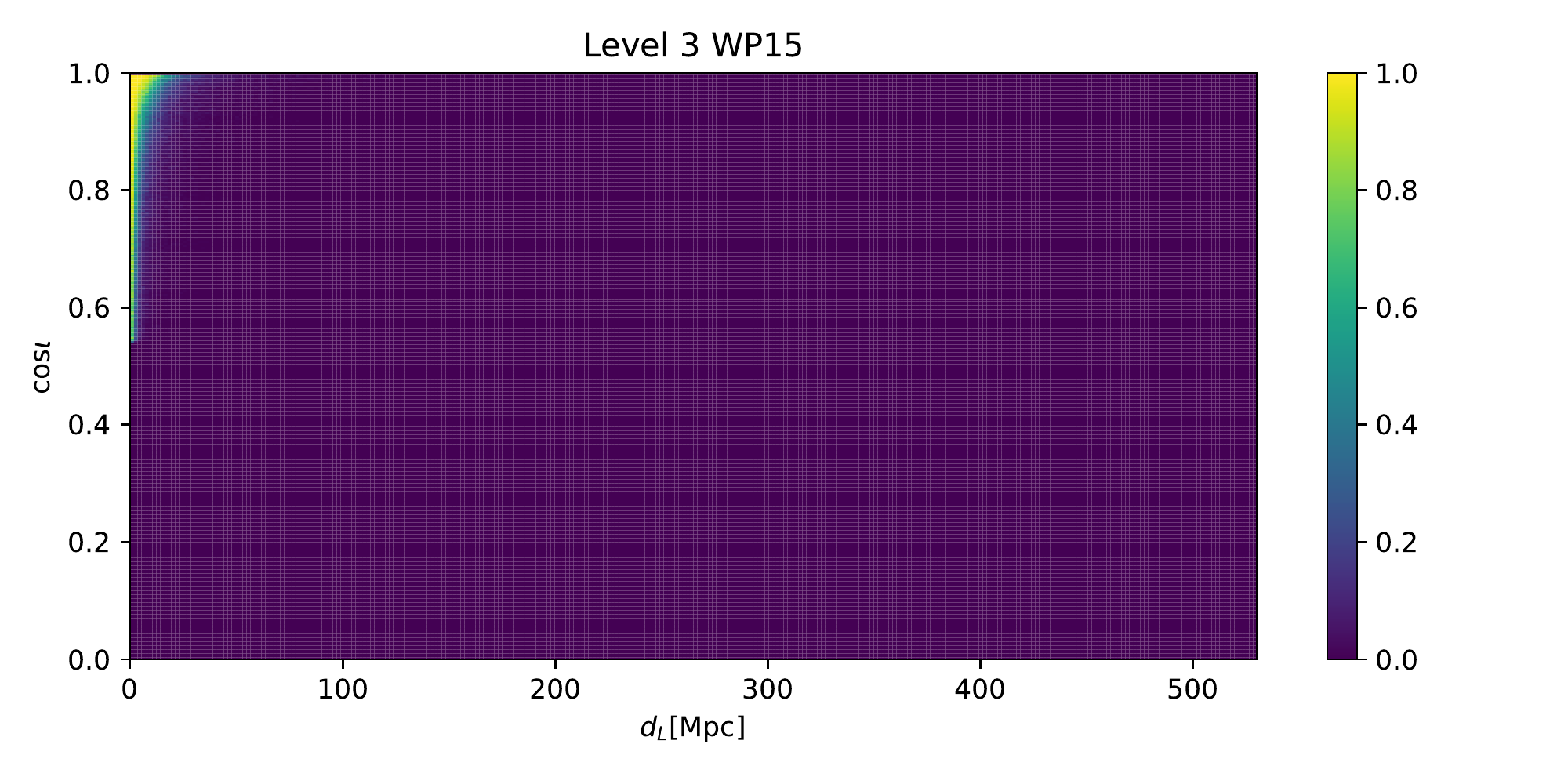}}  
    \end{tabular}
    \caption{Probabilities of detection for the electromagnetic counterparts considered in the study: $p_{{\rm det}}^z$ (top), $p_{{\rm det}}^{z+{\rm AG}}$ (bottom-left) and $p_{{\rm det}}^{z+{\rm AG}+{\rm VLBI}}$ (bottom-right) under the two different hypotheses for the jet's kinetic energy distribution denoted G16 and WP15. See Sect.~\ref{sec:3.2.2} for details on these hypotheses.}
    \label{fig:pdetEM}
\end{figure*}

As for the other electromagnetic counterparts, we determine the probability of detecting the remnant proper motion and making the corresponding angle measurement in addition to the other measurements denoted $p_{{\rm det}}^{z+{\rm AG}+{\rm VLBI}}$ and define $p^{{\rm L}_3} = p_{{\rm det}}^{\rm GW} \times p_{{\rm det}}^{z+{\rm AG}+{\rm VLBI}}$. This is plotted in Fig.~\ref{fig:pdetEM}, where one can see further suppression of events.

\subsection{Selection effects in multimessenger cosmology}

In Appendix~\ref{sec:appendix}, we make estimates of the selection effects impacting the measurements of $H_0$ with GW and electromagnetic data.

In the circumstances of measuring $H_0$ with data from GW170817, that is, a Level 3 scenario during the O2 run, the selection effects are less than 2.0\% over the 60-80~km/s/Mpc range, according to Fig.~\ref{fig:selection}. This is well below the 14\% precision claimed by studies making this measurement \citep{2019NatAs...3..940H}. We conclude that no selection effects significantly impacted the measurement of $H_0$ with the GW170817 data.

However, the selection effect is not negligible compared to the 4\% precision required to resolve the $H_0$ tension \citep{2017NatAs...1E.169F,FMD2018}, particularly with events detected in O3- and O4-type runs, where selection effects reach 2\% and 4\% in the 60-80~km/s/Mpc range, respectively. Then, careful consideration of selection effects becomes necessary.

\section{Prospects of joint electromagnetic and gravitational-wave detections}
\label{sec:4}

In this section, we use the detection probability models to generate a population of events detected jointly in GW and the different electromagnetic counterparts. This allows us to study the dominant effects in the multimessenger detection process.

\subsection{Simulation description}

To evaluate the multimessenger capability to measure $H_0$, we simulate BNS events in a Universe described by flat $\Lambda$CDM cosmology and with $H_0=70$~\hu and $\Omega_m=0.308$.
We generate 80000 BNS mergers uniformly distributed in the sky up to a luminosity distance of $\sim 1.5 d^{r}_{0.2\%}$ for a given detector network with  BNS 0.2\% response distance $d^{r}_{0.2\%}$ (see Fig.\ref{fig:pdetGW}).
We assume a system formation rate uniform in redshift and the BNS masses are generated from a Gaussian distribution with a mean of $1.35 M_\odot$ and standard deviation $0.15 M_\odot$ \citep{2019ApJ...876...18F}, $\cos \iota$ is generated uniform on the unitary sphere and the GW polarization angle is distributed uniformly in the range $\in[0,\pi]$.

For each BNS merger we calculate the measured GW S/N as described in Sect.~\ref{sec:3A} by drawing a measured $d_L$ and $\cos \iota$ from the CF approximation. 
If the GW detected S/N exceeds a threshold of $14$, then we assume that the event has been detected by the HLV network.
Each GW event detected is passed to the electromagnetic follow-up chain described in Sect.~\ref{sec:3B}. 
We use the different \pdetEM as described previously to decide which electromagnetic counterpart is detected according to the value of $d_L$ and $\cos \iota$.
At the end of the simulation, each detected GW event is associated with a flag describing the corresponding scenario: either not detected in the electromagnetic domain or detected with Levels 1, 2, or 3. We repeat this process for O2-, O3- and O4-type GW sensitivities and for the G16 and WP15 electromagnetic counterpart population models, for a total of six GW and population model combinations.

\subsection{Rates of electromagnetic counterpart detections}

In Table~\ref{tab:det_prob_tables}, we show the fraction of GW detections with different electromagnetic information levels predicted for O2-, O3- and O4-like runs. As the population models G16 and WP15 are extremes in terms of afterglow luminosities, one can consider the figures in Table~\ref{tab:det_prob_tables} as confidence interval bounds for the corresponding fractions\footnote{We note that the results on afterglow counterpart fractions presented in this section are consistent with the predictions of \cite{DDM2019}, though the treatment of the GW and jet motion in the present study is more refined.}.

In an O2-type run, 52\% of the GW events are expected to have a detectable kilonova counterpart. Since we assumed the optical instruments can cover a fraction $p_{\rm night} = 52 \%$ of the sky, this means that the kilonova magnitude was not a limiting effect for multimessenger events during O2. 
In other words, the GW detection probability dominates the multimessenger detection probability, as was predictable from Figs.~\ref{fig:pdetGW} and \ref{fig:pdetEM}.

Conversely, the fraction of events expected to have been detected in O2 with Levels 2 and 3 are between 4\% and 12\% and between 1\% and 7\%, respectively. This is from 10\% to 25\% and from 2\% to 15\% of kilonovae with detectable afterglow light curve and remnant proper motion, respectively. Therefore, as of O2, the sample of events with afterglow counterpart was limited by selection.
As it can be seen from Figs.~\ref{fig:pdetGW} and \ref{fig:pdetEM}, this is mostly because electromagnetic signals are much more sensitive to binary inclination than GW signals.

GW170817 was a Level 3 event during O2 \citep{2017ApJ...848L..12A,2019NatAs...3..940H}. In light of our study, the probability of this to occur was between 1\% and 7\%, and GW170817 was a very lucky event. 

In O3-type runs, the kilonova counterparts are still largely in reach of follow-up instruments. 
However, the probability of measuring $\cos \iota$ with afterglow counterparts is strongly reduced compared to an O2-type run, meaning the multimessenger detection process is strongly dominated by electromagnetic selection.
Nonetheless, there is a non-negligible probability of an afterglow counterpart to a GW event.
In the first 6 months of O3, GW190425 \citep{2020ApJ...892L...3A} was the only plausible BNS event observed. For this event, there was no electromagnetic counterpart reported by large field-of-view facilities \citep{2020MNRAS.492..863C}. 
Let us note that, as opposed to the events considered in this simulation, GW190425 was detected by only two detectors (LIGO Livingston and Virgo), thus producing a large sky-localization area covering 10,200 deg$^2$.

From our simulations, we estimate that maintaining an O2-like fraction of radio-detectable afterglows during an O3-like GW run would require a factor of five increase in the radio sensitivity. That is, a $F_{\rm lim}$ on the order of $3~\mu$Jy, as projected for the Square Kilometer Array 1 ``Mid'' band \citep{DHSL2009}.

In an O4-like run, GW detectors detect a large fraction of BNS mergers at higher distances. In this case, the effects of magnitude limitations start to kick in, with only half of the kilonovae with sufficient flux. Taking into account sky-position limitations, this number is one out of 4. In O4-type runs, it should prove extremely rare to obtain an electromagnetic $\cos \iota$ measurement (Level 2 and Level 3). 

\begin{figure}
    \centering
    \includegraphics[scale=0.51]{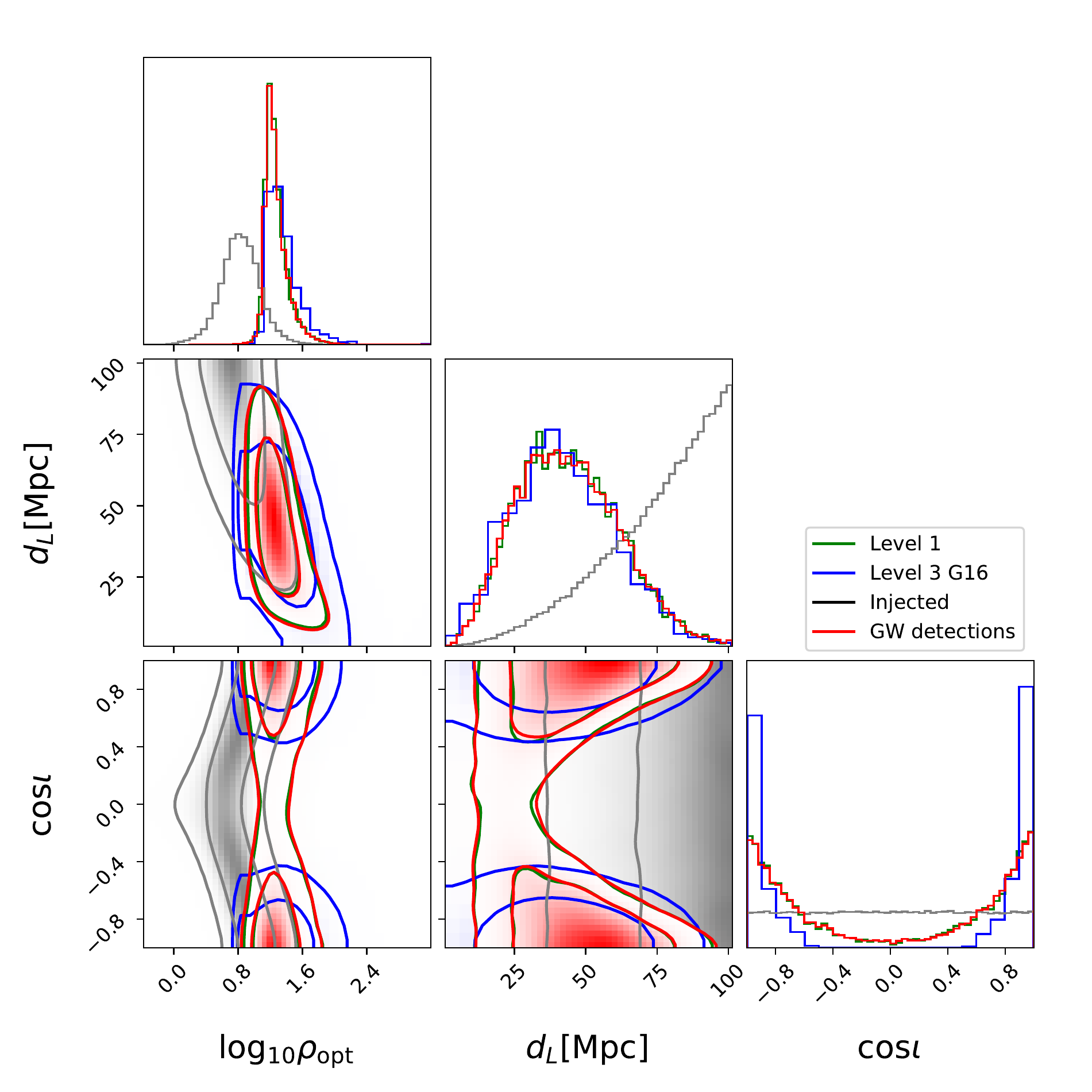}
    \caption{Histograms for the distributions of inclination angle $\iota$, luminosity distance $d_L$ and optimal-filter GW signal-to-noise ratio (see Eq.~\ref{eq:SNR}) for all simulated BNS mergers (black line), those detected with GW only (red), and those detected with Level 1 (green line) or Level 3 (blue line) for an O2-type run.}
    \label{fig:O2_g16_vlbi}
\end{figure}

\begin{figure}
    \centering
    \includegraphics[scale=0.51]{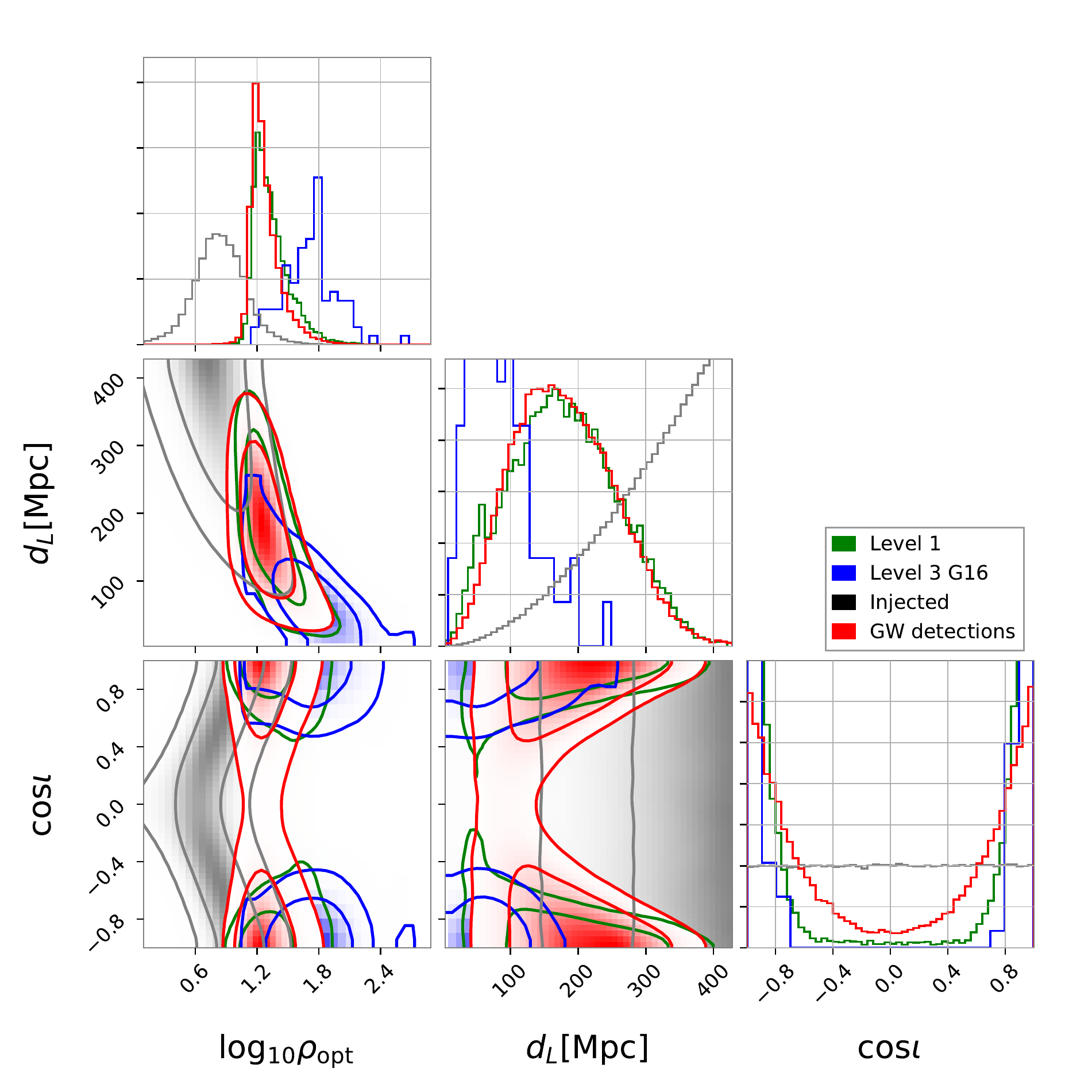}
    \caption{Same as Fig.~\ref{fig:O2_g16_vlbi}, for an O4-type run.}
    \label{fig:O4_g16_vlbi}
\end{figure}

The different selection biases introduced by the GW and electromagnetic detections processes when going from an O2-like to an O4-like run are depicted in Figs.~\ref{fig:O2_g16_vlbi}-\ref{fig:O4_g16_vlbi}.
The two figures show the distributions in $d_L$, $\cos \iota$, and optimal-filter S/N detected in different levels.
In an O2-like run, the GW and electromagnetic detections roughly probe the same BNS mergers, with the exception of some regions in $\cos \iota$ that forbid a detection by VLBI. 
On the other hand, in an O4-like run, the electromagnetic and GW detection clearly corresponds to a different subpopulation of sources. In fact, the electromagnetic facilities can observe only close-by events--which thus have a high GW S/N--corresponding to a very small fraction of GW-detected events.

In Table~\ref{tab:det_rate_tables} we report the average number of years of observation required to detect BNS mergers with different levels of EM counterparts. We assume a BNS merger rate of $R_0=320^{+490}_{-240} \rm{Gpc^{-3}\, yr{^{-1}}}$ as estimated after the first six months of O3 \citep{2020arXiv201014533T}. We note that this is overestimated as this is based on a conservative choice for the detection threshold. For this simulation we selected confident BNS detections with S/N~$> 14$ ($\sim 8$ in each detector) while current GW detectors can detect BNS with S/N $\sim 8$ \citep{2018LRR....21....3A}. Choosing a detection threshold of S/N $\sim 8$ corresponds to a surveyed volume $\sim 6$ times larger, and thus an observation time to detection reduced by the same factor. Table~$\ref{tab:det_prob_tables}$ reports the relative fractions of events detected with different levels of EM emission. 
From the results in Table~\ref{tab:det_rate_tables} we conclude that the probability of detecting a BNS merger during a one-year O2-like run is small, and especially so for a loud event such as GW170817.


\begin{table*}[t!]
\caption{Average fraction of GW events observed with different electromagnetic counterpart levels.}
\begin{center}
\begin{tabular}{|>{\centering}p{30mm }|>{\centering}p{20mm}|>{\centering}p{20mm}|>{\centering}p{20mm}|>{\centering}p{20mm}|>{\centering}p{30mm}|}

\hline

& \multicolumn{5}{c|}{Electromagnetic information level} \\

GW Run & Level 1 & \multicolumn{2}{c|}{Level 2}
& \multicolumn{2}{c|}{Level 3} \\

&  & WP15 & G16
& WP15 & \multicolumn{1}{c|}{G16} \\

\hline \hline

O2-like & 52\%  & 4\% & 12\%
& 0.67\% & \multicolumn{1}{c|}{7\%} \\

\hline 

O3-like & 45\%  & 1.56\% & 6.13\%
& 0.18\% & \multicolumn{1}{c|}{1.70\%} \\

\hline 

O4-like & 26\%  & 0.37\% & 3.50\%
& 0.01\% & \multicolumn{1}{c|}{0.25\%} \\

\hline 

\botrule

\end{tabular}
\end{center}
\label{tab:det_prob_tables}
\end{table*}

\begin{table*}[t!]
\begin{center}
\caption{Average observation time (years) needed to detect one BNS merger with different EM counterpart levels. Observation times are computed assuming $R_0=320^{+490}_{-240} \rm{Gpc^{-3}\, yr{^{-1}}}$, the best estimate after the first half of the O3 run. We assume a network S/N of at least 14 for GW detection; relaxing to a threshold of 8 would decrease these waiting times by a factor of $\sim$ 6 (see text for details).}
\label{tab:det_rate_tables}
\begin{tabular}{|>{\centering}p{30mm }|>{\centering}p{20mm}|>{\centering}p{20mm}|>{\centering}p{20mm}|>{\centering}p{20mm}|>{\centering}p{30mm}|}

\hline

& \multicolumn{5}{c|}{Electromagnetic information level} \\

GW Run & Level 1 & \multicolumn{2}{c|}{Level 2}
& \multicolumn{2}{c|}{Level 3} \\

&  & WP15 & G16
& WP15 & \multicolumn{1}{c|}{G16} \\

\hline \hline

O2-like & $15^{+47}_{-9}$ & $200^{+617}_{-124}$ & $68^{+205}_{-42}$
& $1200^{+3600}_{-740}$ & \multicolumn{1}{c|}{$117^{+350}_{-72}$} \\

\hline

O3-like & $2^{+6}_{-1}$  & $62^{+187}_{-38}$ & $16^{+47}_{-10}$
& $540^{+1610}_{-320}$ & \multicolumn{1}{c|}{$57^{+171}_{-35}$} \\

\hline 

O4-like & $0.5^{+1.4}_{-0.3}$  & $33^{+99}_{-20}$ & $3^{+10}_{-2}$
& $1220^{+3677}_{-741}$ & \multicolumn{1}{c|}{$49^{+147}_{-30}$} \\
\hline 
\botrule
\end{tabular}
\end{center}
\end{table*}

\section{Forecast on multimessenger $H_0$ measurements}
\label{sec:5}

\subsection{Method}
We now set to quantify the benefit of electromagnetic-provided information in measuring $H_0$. We use the statistical framework described in Appendix~\ref{sec:appendix} with the posterior as in Eq.~\ref{eq:mandel}. The likelihood $p(\textbf{d}|H_0)$  encodes the statistical uncertainties of the GW and electromagnetic data $x_{\rm GW/EM}$ measurements:
\begin{multline}
    p(\textbf{d}|H_0) = \int \d z \, \d \cos \iota \, p_{\rm GW}(x_{\rm GW}|d_L(H_0,z), \cos \iota) \\
    \times p_{\rm EM}(x_{\rm EM}|\cos \iota, z) p_{pop}(z, \cos \iota|H_0),
    \label{eq:likeli}
\end{multline}
where $p_{pop}(z, \cos \iota|H_0)$ is the assumed population distribution in redshift and inclination for the entire BNS population.

The function $p_{\rm GW}(x_{\rm GW}|d_L(H_0,z),\cos \iota)$ is the GW likelihood, which provides the GW-detected source parameters distribution for given true parameters. We take the CF approximation for consistency with our earlier computation of selection effects.

Likewise, $p_{\rm EM}(x_{\rm EM}|\cos \iota,z)$ is the likelihood for electromagnetic measurement of source parameters $\cos \iota$ and $z$.
This is informative on $z$ only for a Level 1 scenario and on both $z$ and $\cos \iota$ for higher-level scenarios. It is not informative on $d_L$.
We decompose this electromagnetic likelihood as $p_{\rm EM}(x_{\rm EM} | z)  p_{\rm EM}(x_{\rm EM}| \cos \iota)$ by supposing the electromagnetic measurements of redshift and angle are independent. This is reasonable as the redshift information is deduced from the host galaxy alone while the $\iota$ information is expected to be provided by the jet itself.

We assume the redshift measurement is unbiased and set $p_{\rm EM}(\hat x_{\rm EM}|z)$ to a Gaussian distribution centered on the true event redshift with standard deviation $5 \cdot 10^{-4}$ for all our scenarios. This is the same accuracy measured for GW170817's redshift \citep{2017Natur.551...85A}. For Level 1 scenarios, the electromagnetic counterpart is uninformative on $\cos \iota$ and we set $p_{\rm EM}(x_{\rm EM}|\cos \iota)$ to a flat function.

For Level 2 scenarios, we assumed that one can obtain an unbiased $\iota$ constraint at the level of that deduced from GW170817's afterglow light curve alone. That is, a Gaussian constraint with a $12 \deg$ 1-$\sigma$ uncertainty \citep{TVRRB+2019,HMAKB+2019} for $p_{\rm EM}(\hat x_{\rm EM}|\cos \iota)$. 

For Level 3 scenarios, we based our predicted constraints on those of GW170817 and set $p_{\rm EM}(\hat x_{\rm EM}|\cos \iota)$ to an unbiased Gaussian constraint with a width of $4 \deg$ \citep{MDGNH+2018, GSPGY+2019, 2019NatAs...3..940H} for all the events. 
We discuss the validity and impacts of these assumptions in Sect.~\ref{sec:6}.

Using Eqs.~\ref{eq:mandel}, \ref{eq:explicit} and \ref{eq:likeli}, we simulated the $H_0$ measurement process of 500 binary systems in all three observing scenarios and all three GW sensitivity hypotheses. For the individual measurements, we systematically assumed a prior on $H_0$ uniform in [40, 120]~\hu.

We then combined the measurements of the first 100 events to emulate a thread of multimessenger events. We repeated the combining step after reordering the 500 events to reproduce different possible time orderings of the events. This allowed us to study the reconstruction of $H_0$ by the multimessenger measurements, and in particular the speed of convergence.

\subsection{Bulk comparison of observing scenarios}
\label{sec:5A}

Breaking the $d_l-\cos \iota$ degeneracy is fundamental for measuring $H_0$. Let us show this by inferring $H_0$ using \textit{only} Level 2 or 3 scenarios in comparison with Level 1 scenarios.

\begin{figure}
    \centering
    \includegraphics[width=\linewidth]{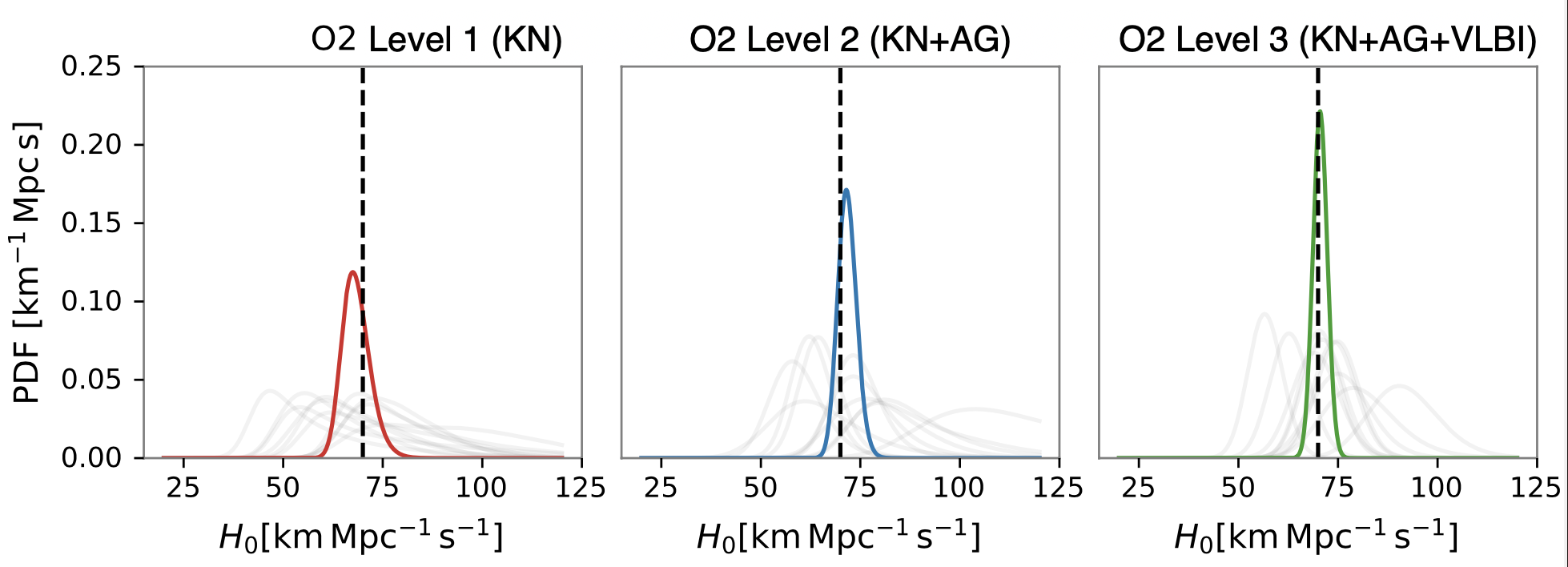}
    \caption{Posterior on $H_0$ obtained for ten BNS events observed with different level of electromagnetic information on $\iota$ in an O2-type run. \textit{Left}: Level 1 (no electromagnetic information on $\iota$), \textit{Center}: Level 2 (electromagnetic-based $\iota$ precision of 12 deg), \textit{Right}: Level 3 (electromagnetic-based $\iota$ precision of 4 deg).}
    \label{fig:comparison_O2_toy}
\end{figure}

Fig.~\ref{fig:comparison_O2_toy} shows the $H_0$ posteriors obtained by combining 10 BNS events in different observing scenarios.
From the plot, we can see that, when the knowledge of $\cos \iota$ is refined by electromagnetic observations, the estimation of $H_0$ improves, as also noted in \cite{2019PhRvX...9c1028C}. We can also observe that the $H_0$ posterior reaches Gaussian convergence after under 10 events.

\begin{figure}
    \centering
    \includegraphics[width=\linewidth]{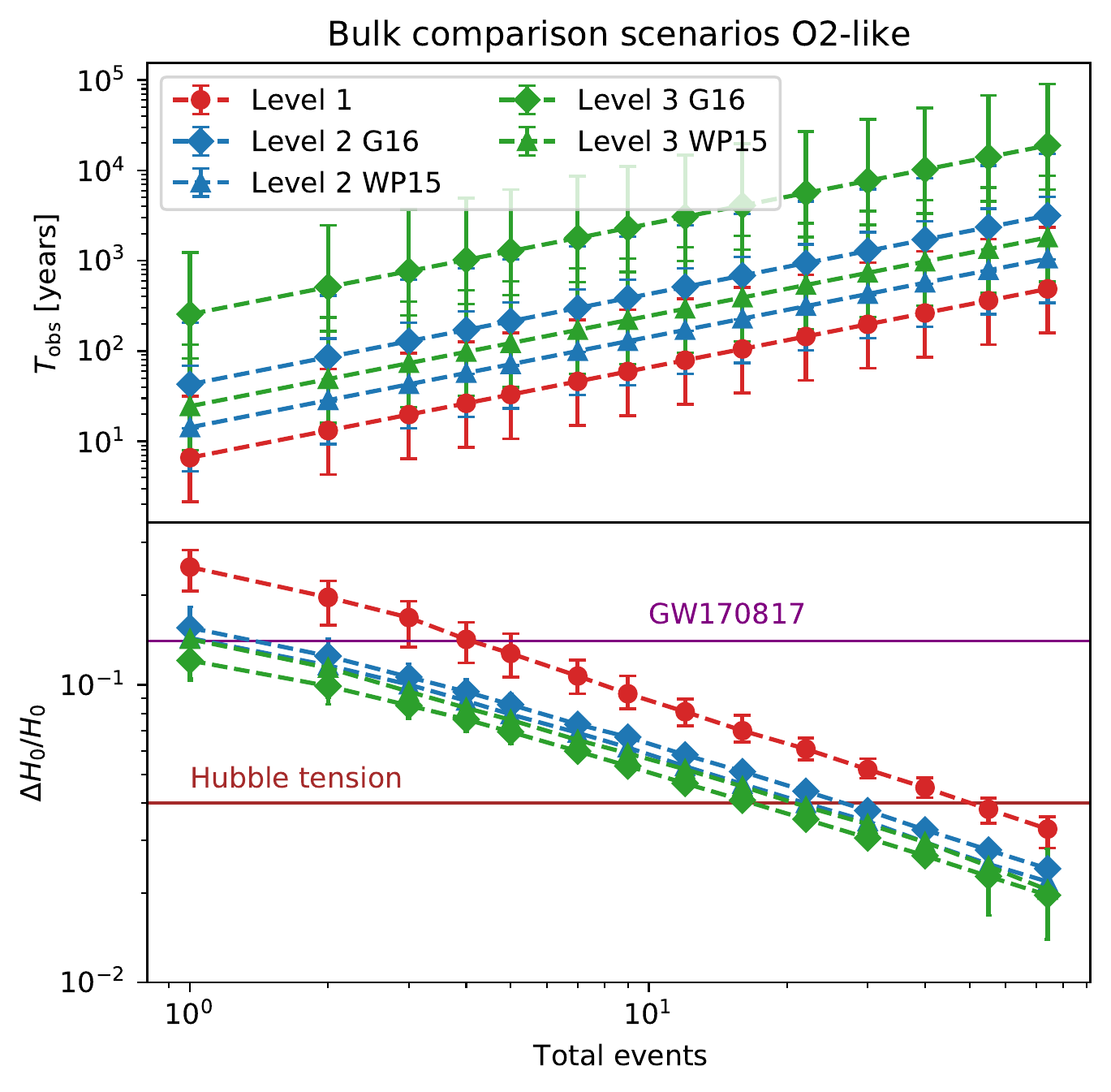}
    \caption{Convergence of $H_0$ with number of detected events. \textit{Bottom}: Precision on the estimation of $H_0$ with 68.3\% confidence intervals as a function of the number of events detected at each EM level. Here, all events are assumed to be at a given electromagnetic information level, as denoted in different colors. We also indicate the tension in the Hubble constant and the precision on $H_0$ obtained with GW170817 and counterparts. \textit{Top}: Expected number of years of continuous observation required to detect events in various multimessenger scenarios.}
    \label{fig:toy_scenario}
\end{figure}

The bottom panel of Fig.~\ref{fig:toy_scenario} shows the relative uncertainty $\Delta H_0 / H_0$ with $1-\sigma$ confidence intervals for the $H_0$ estimation as more events are detected. 
The uncertainty corresponds to the different population realizations of the detected events. In Fig.~\ref{fig:toy_scenario}, one can read that a single Level 3 event during an O2-type run results on average in the uncertainty of $14\%$ on the estimation of the Hubble constant, as observed for GW170817 \citep{2019NatAs...3..940H}.

It is clear that \textit{(i)} the precision on $H_0$ improves as more events are combined, \textit{(ii)} the convergence is faster when $\cos \iota$ is more constrained from the electromagnetic emission, \textit{(iii)} above about ten events \citep{2018Natur.562..545C}, the combined $H_0$ posterior becomes Gaussian, with $\Delta H_0 / H_0 \sim \Theta/\sqrt{N}$ where $N$ is the number of events, allowing us to define $\Theta$ as an effective single-event $H_0$ estimation standard deviation.

In the top panel of Fig.~\ref{fig:toy_scenario}, we show the expected number of years of continuous observing required to detect the number of events read on the bottom horizontal axis. We assume an O2-like sensitivity\footnote{We consider an O2-like sensitivity for Fig.~\ref{fig:toy_scenario} for computational reasons. Using O3- or O4-like sensitivities decreases the number of Level 3 events in our simulation to the point that emulating the $H_0$ measurement with even ten such events was computationally infeasible.} and the detection rates reported in Table~\ref{tab:det_rate_tables}.
From the bottom panel of Fig.~\ref{fig:toy_scenario}, the number of Level 2 events required to resolve the $H_0$ tension assuming an O2-like sensitivity is $\sim 25$. According to the top panel, one would need 100 to 600 years of observation to collect these events, depending on the population model.

Counterparts to GW events detected with different GW sensitivities probe different regions of the distance and inclination parameter space (compare Figs.~\ref{fig:O2_g16_vlbi} and \ref{fig:O4_g16_vlbi}). Therefore, one cannot deduce the number of events or observing time required to resolve the $H_0$ tension for an O4-type run from those for an O2-type run by simply comparing the GW detection rates. In the next section, we derive the number of events required to resolve the $H_0$ tension assuming a long-lived O4-type run.

\begin{figure}
    \centering
    \includegraphics[width=\linewidth]{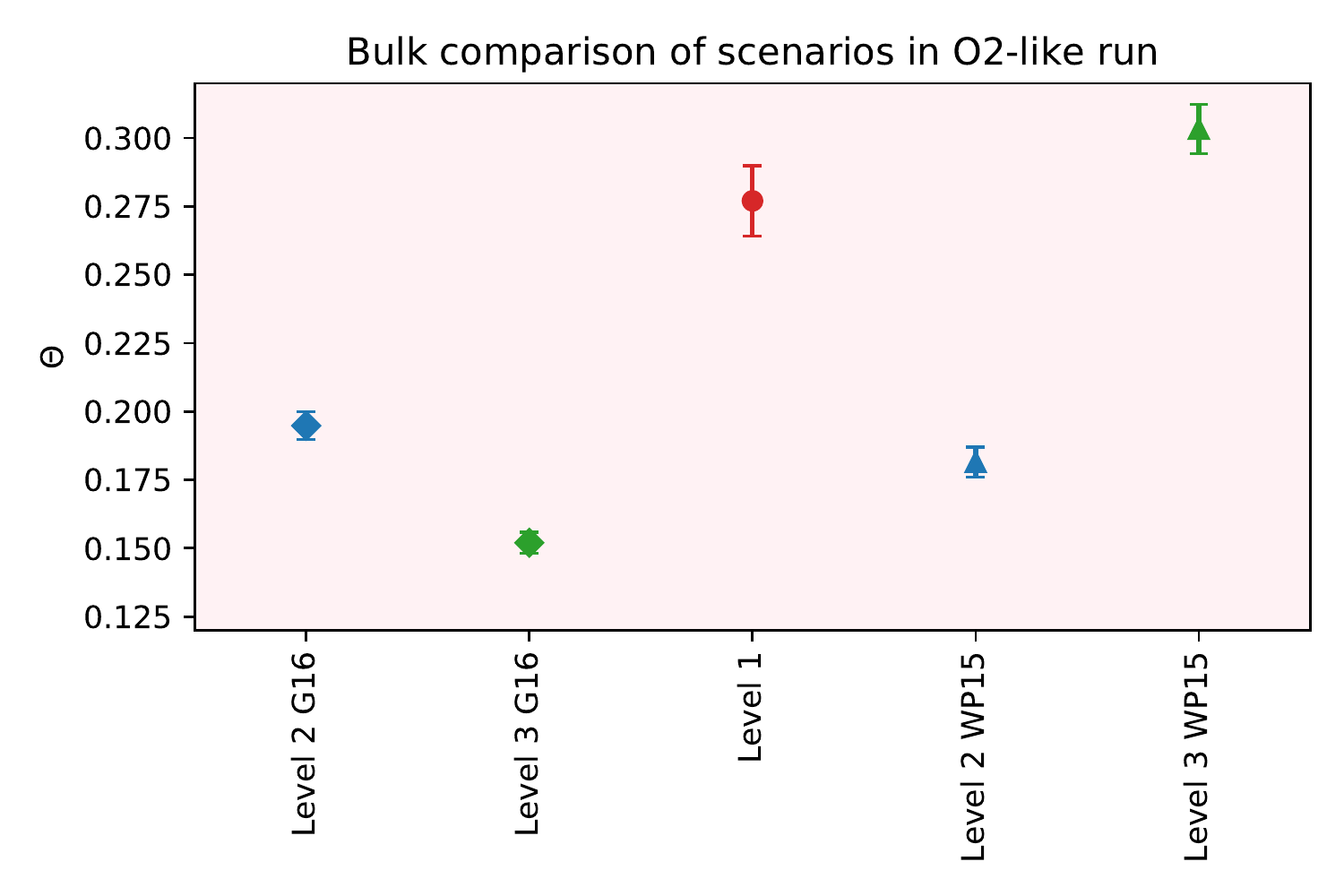}
    \caption{Effective single-event standard deviation $\Theta$ in different observing scenarios, assuming all events are at a given electromagnetic information level, in an O2-type GW run.}
    \label{fig:improvement_toy}
\end{figure}

In Fig.~\ref{fig:improvement_toy} we show the values of the average effective single-event standard deviation $\Theta$ of the different scenarios, as fit on the curves in Fig~\ref{fig:toy_scenario}. There is a clear boost in the $H_0$ convergence speed when considering the information in $\cos \iota$ from the electromagnetic counterparts. We find that a $\cos \iota$ precision of $4 \deg$ (Level 3) provides a convergence that is 1.4 times faster than one of $12 \deg$ (Level 2), which is itself about 1.5 times faster when there is no angle information at all from the electromagnetic domain. Roughly, it means that the $H_0$ accuracy reached combining 10 Level 2 events is equivalent to that reached by combining 5 Level 3 events.

We also find that detectors with better sensitivities will be able to better constrain $H_0$, even without $\cos \iota$ measurements from electromagnetic counterpart. This is due to the higher redshifts of the events used to infer $H_0$, as we consider a constant uncertainty in redshift measurements.
Considering events without electromagnetic contributions to $\cos \iota$ for O4-type runs, we find an average effective single-event standard deviation of 14\%, consistent with previous simulations \citep{2018Natur.562..545C,2020PhRvD.101l2001G}.

We note that for a given GW sensitivity, there is no difference in this first approach between G16 and WP15 population models as these only impact the probabilities of detecting the electromagnetic counterparts. This is clear in Figs.~\ref{fig:toy_scenario} and \ref{fig:improvement_toy}.

\subsection{Considering realistic detection rates}

In a real observing run, not all the GW events with a redshift estimation will have $\cos \iota$ measurements from electromagnetic counterparts. 
In a second approach, we estimate the $H_0$ convergence by including the relative detection rates of the different electromagnetic counterparts. 
More precisely, we generated threads of events as in the bulk comparison of the previous section, but successively allowing for only Level 1 events, then up to Level 2, then up to Level 3, to quantify the acceleration of $H_0$ convergence each EM level allows. 

We then simulated the $H_0$ measurement for each of them, and combined their measurements throughout the first 100 events. We repeat this process 500 times with varying event time orderings to simulate different realizations of the subset of 100 detections.

\begin{figure*}
    \centering
    \includegraphics[width=\linewidth]{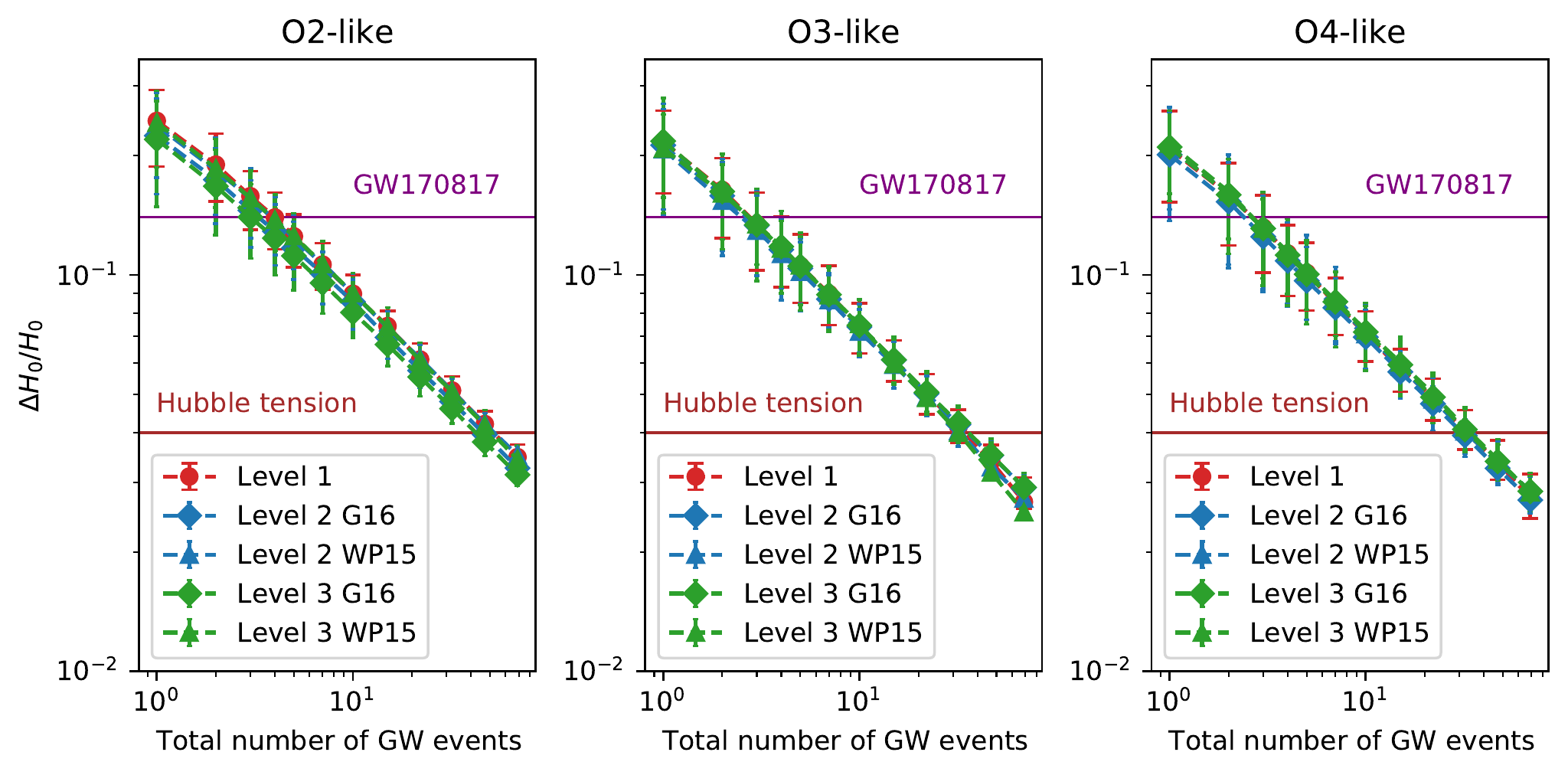}
    \caption{Same as Fig.~\ref{fig:toy_scenario} (bottom panel), with the bottom axis counting the total number of GW events, regardless of the nature of the EM counterpart. Therefore, this considers the realistic detection rate of events with different electromagnetic counterparts.}
    \label{fig:real_scenario}
\end{figure*}

\begin{figure}
    \centering
    \includegraphics[width=\linewidth]{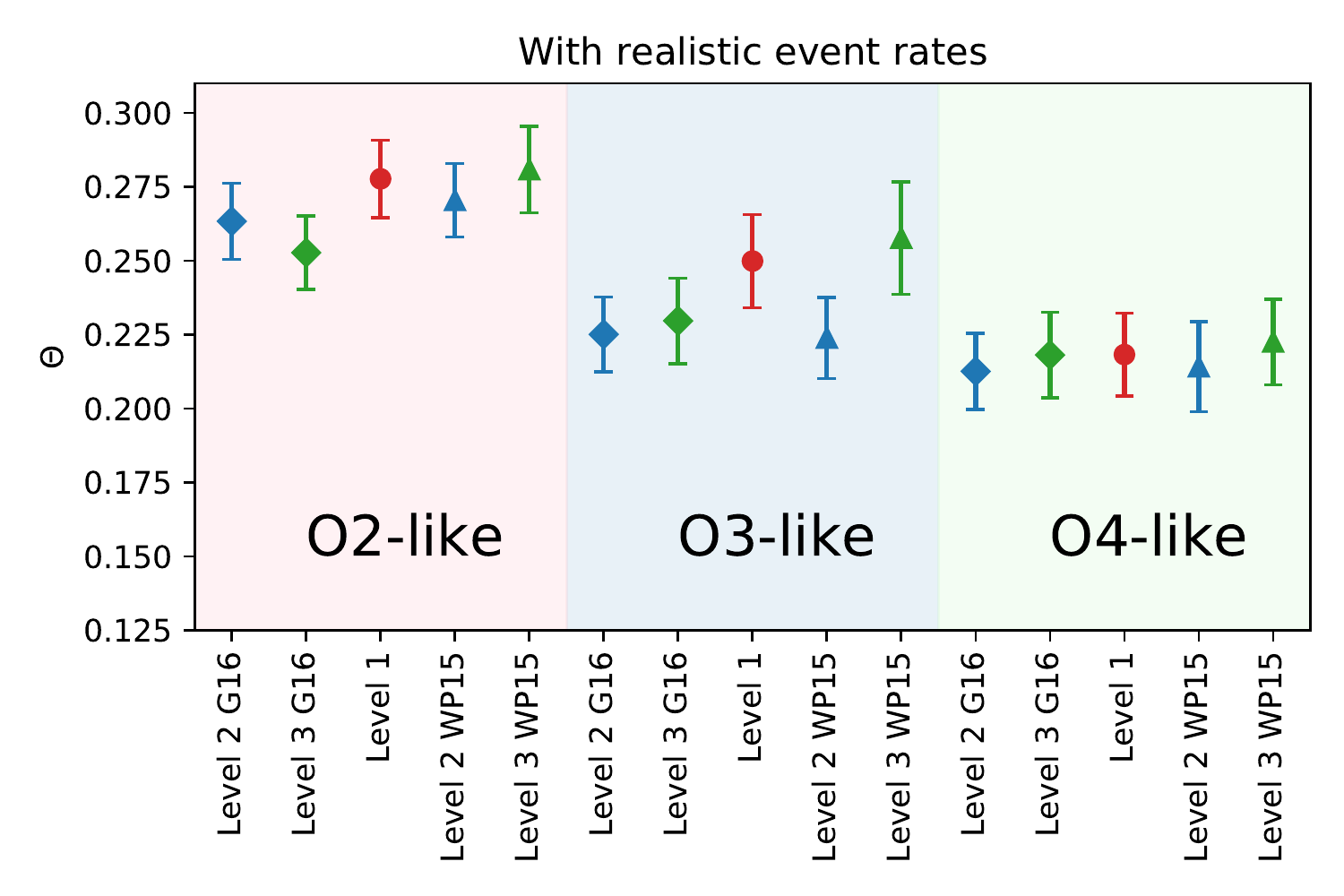}
    \caption{Same as Fig.~\ref{fig:improvement_toy}, but assuming a realistic rate of electromagnetic counterpart detection.}
    \label{fig:improvement_real}
\end{figure}

Fig.~\ref{fig:real_scenario} shows the $H_0$ convergence as a function of the total number of detections, while Fig.~\ref{fig:improvement_real} shows the single-event standard deviation $\Theta$ for all scenarios. According to Table~\ref{tab:det_prob_tables}, about 7\%--23\% of O2 events with associated redshift would have had Level 2 information on $\iota$, and about 1\%--13\% Level 3. As seen in Fig.~\ref{fig:real_scenario}, this detection fraction is enough to somewhat improve the $H_0$ convergence. In other words, the convergence speed allowing for Level 2 and 3 events is larger than with kilonova events alone.

The situation changes drastically when we consider the multimessenger events in O3- and O4-type observing runs. In this case, the fraction of Level 2 or Level 3 events are so small that on average they bring no additional improvement. Indeed, the precision on $H_0$ obtained with a given number of GW events does not change whether we allow for afterglow counterparts or if we do not.
In particular, for an O3-like run, only allowing for Level 3 events and assuming the optimistic G16 population prescription could provide a slight acceleration in the $H_0$ narrowing-down, while for all the other cases the improvement is negligible. In O4-type runs, neither Level 2 or Level 3 events should statistically speed up the convergence of $H_0$.

During an O4-type run, Level 2 or 3 events are too rare to significantly improve the $H_0$ convergence and shorten the time needed resolve the $H_0$ tension. Relying on Level 1 events only is just as fast. In Sect.~\ref{sec:6}, we argue that, in this case, discarding measurements of the inclination angles from afterglows prevent further biases in the $H_0$ estimate.

For O4-like runs, Table~\ref{tab:det_rate_tables} shows that one Level 1 events are detected every 6 months on average. Fig.~\ref{fig:real_scenario} shows that, to resolve the $H_0$  tension problem, 30 such events are required. Thus, with an O4-like sensitivity, 15 years of data taking are necessary to collect the number of detections with measured redshift.

\section{Discussion}
\label{sec:6}

In this article, we studied the prospects of measuring the Hubble constant with GW standard sirens coupled to inclination angle measurements from merger afterglow counterparts. We first studied the potential impact of selection effects in multimessenger cosmology and showed that these were negligible in the $H_0$ measurement reported after GW170817. We illustrated how selection effects increase with the GW sensitivity. For events in future GW observing runs, selection effects will be important and should be taken into account.

We then studied the likelihood of detecting the electromagnetic counterparts required for multimessenger measurements of $H_0$. We showed that for O2- and O3-type GW sensitivities, their optical magnitudes are not limiting in the detection of kilonova signals, and therefore in acquiring source redshifts. This is true provided GW interferometers are operating and follow-up facilities available and efficient.

We showed that, in O4-like runs and beyond, the GW detection probability will extend further in distance compared to the electromagnetic detection probabilities, largely decreasing the likelihood of detecting electromagnetic counterparts.This will especially be the case when second-generation GW interferometers reach their design sensitivity level. Fortunately, the lower number of observable counterparts should be counter-balanced by the advent of large field-of-view, high-cadence optical instruments such as the ZTF and LSST, which are not limited in terms of skymap coverage (but only by their limiting magnitude). Also, the advent of additional GW interferometers such as KAGRA \citep{KAAAA+2019} will largely improve the median GW localization skymap down to $\sim 40~\rm{deg}^2$ \citep{2018LRR....21....3A}, paving the way to more effective follow-up by smaller field-of-view instruments and therefore better-sampled kilonova light curves and better leveraging of events.

In the future, third-generation GW detectors such as the Einstein Telescope \citep[ET,][]{ET} and Cosmic Explorer \citep{CE} will open up the detection range of even larger redshifts. Naturally, the low-redshift events will still be accessible, with a much better sky resolution that can be  covered entirely by small field-of-view facilities. For  larger redshifts, follow-up observations will be limited by the sky localization, with an average resolution of $200~{\rm deg}^2$ for BNS at $z=0.1$ with ET. This coverage issue combined with the unprecedented dimness of the counterparts at these redshifts may call for totally different follow-up strategies for the high-cadence large field-of-view survey facilities. Supposing the source is identified, photometric and spectroscopic follow-up would still be limited to $z = 0.5$ and $z=0.3$, respectively, for the largest optical telescopes such as the Extremely Large Telescope \citep{2020JCAP...03..050M}. Therefore, the access to those events at cosmic redshifts should rely on the observation of counterparts other than kilonova, such as the short gamma-ray burst, the detection of which could be facilitated by the GW signal being present in the ET band hours before merger.

We finally studied whether the observation of merger afterglow signals and subsequent measurements of $\cos \iota$ will significantly accelerate the narrowing-down of $H_0$ when combined with GW detections in the future. We considered only the afterglow signal--its photometry and imaging--as potential providers of $\iota$ measurements. We deemed the other counterparts such as the kilonova and gamma-ray burst unfit for such a measurement, for their still large modeling uncertainties.

We considered an optimistic measurement model in which all events with an afterglow counterpart contribute a $\iota$ measurement with an accuracy comparable to GW1710817, for both afterglow photometry and imaging. This is an optimistic assumption as the uncertainty on $\iota$ depends on the number of photometric points detected from the light curve, and thereby on the event distances and density of the follow-up. Also, only a subset of the follow-up campaigns is expected to provide such detailed multiwavelength photometric data. Taking the variability of the follow-up scope and data quality into account is a possible extension to this study.

Furthermore, our analysis assumes that $\cos \iota$ estimated from EM is accurate, which is a simplifying assumption. For both the afterglow photometry and imaging analysis, the leading uncertainty in the electromagnetic modeling is the treatment of the jet lateral expansion \citep{GSPGY+2019}. The jet expansion affects both the time of afterglow peak flux \citep[e.g.,][]{DDM2019} and the dynamics of the VLBI image \citep{2021arXiv210105138F}, possibly biasing and widening the electromagnetic posterior on $\iota$ more than assumed here.

From the point of view of observations, most of the indeterminations in measuring $\iota$ from GW170817's afterglow lie in the very early phases of the light curve and the late phases of the source displacement curve \citep[see the posterior sampling in][]{GSPGY+2019}. Acquiring early photometric data with deeper searches would have provided a better estimate of $\iota$ from this source. Recently, radio points were acquired from GW170817 to flux levels much deeper than the early radio monitoring of GW170817 \citep{2021arXiv210304821B}, proving that this is in reach of current radio facilities. We thus advocate for such early deep searches.

Also, we assumed perfect kilonova detection and identification above a magnitude threshold over the accessible sky, and our expression of the jet proper motion in Eq.~\ref{eq:ppm} is clearly an overestimate, leading to overpredict the number of sources with detectable proper displacement.

Finally, the effect of peculiar velocities can bias the estimation of $H_0$. Galaxies' peculiar motions can be as high as $v_p=300 ~\rm{km/s}$, with an associated error of their measurement of the same order \citep{Mukherjee:2019qmm}. This corresponds to a redshift correction (and additional uncertainty on $H_0$) of 10\% for events such as GW170817, and it is fundamental to take into account. With future GW detectors, whici will detect BNSs up to redshift $\sim 0.1$, this type of correction will generally be negligible though it will remain important in the case of close-by, high-S/N events.

For all these reasons, our hypotheses are optimistic. Even so, we have found that, for all GW runs after O3, events with afterglow counterparts should prove so rare that, statistically, using the afterglow counterparts when available for the multimessenger measurement of $H_0$ on individual events will not bring any acceleration to the measurement of $H_0$.

We found that, for the electromagnetic measurement of $\cos \iota$ to significantly increase the $H_0$ convergence, the GW and electromagnetic detection probabilities should be comparable, or the understanding of electromagnetic emission from compact binary mergers should drastically improve, to the point where a degree-level precision on $\iota$ is accessible from a typical afterglow light curve. Even then, care should be taken with systematic effects in angle measurements, for example, from VLBI imaging, to not create a second tension on $H_0$.

The current state of kilonova modeling does not allow for robust measurements of the inclination angle. This may change in the future as the model uncertainties will probably reduce after more signals are observed. Given the above-mentioned rates of kilonova associations with GW, we estimate that kilonovae could accelerate the narrowing-down of $H_0$ if a $\sim 10\%$ calibration can be reached between kilonova data (such as light curves or color evolution) and inclination angle. In this respect, the advent of wide field-of-view, high-cadence optical facilities is an asset as they are expected to collect a large sample of kilonovae detected both serendipitously and as GW counterparts. With tens of well-sampled kilonova light curves, the $10 \%$ accuracy level for the light curve versus inclination relation may be within reach, especially if these sources are standardizable \citep{KRA2019,2020NatCo..11.4129C,2020PhRvR...2b2006C}. This perspective would truly allow multimessenger cosmology to develop.

Our results should not be misunderstood: If the opportunity of making an electromagnetic measurement of $\cos \iota$ occurs, then it should obviously be made as the improvement on $H_0$ from such an event is important. All that we found is that, statistically, being able to make such measurements will not significantly speed up the narrowing-down of $H_0$ and the resolution of the Hubble tension on the long run. This is because of the rareness of electromagnetic counterparts and their still too-poor constraints on $\cos \iota$.

We proved that the electromagnetic-provided  $\cos \iota$ measurements will likely not drive the $H_0$ narrowing-down. 
Therefore, direct biases to $H_0$ through electromagnetic mismeasurements of $\cos \iota$ should not be feared. 
However, as the detection probabilities of the electromagnetic counterparts should dominate the selection effect for GW-EM standard sirens, incorrectly modeled dependence of the kilonova signal on the inclination can lead to $H_0$ biases through uncontrolled selection effects, as discussed in \cite{2018Natur.562..545C,2020arXiv200602779C}. Correct modeling of the kilonova signal to control the selection effects in follow-up campaigns should be a point of care for future endeavors in multimessenger cosmology.

Once these selection biases are dealt with, the limiting uncertainty in multimessenger cosmology should be the GW data calibration.
This uncertainty is at the level of $\sim 1\%$ \citep{2016RScI...87k4503K}, below the Hubble constant tension, and therefore should not impede the resolution of the tension by multimessenger cosmology when combining a low number of events. However, a systematic effort on calibration uncertainties when combining a large number of events should be performed.

As the afterglow counterparts should not accelerate the measurement of $H_0$, we can state that the number of multimessenger events necessary to resolve the $H_0$ tension is still that given by \citet{CHMEV+2017} and \citet{MFPWN2019}, that is, 20--50. This represents about fifteen years of continuous O4-level GW observation.

\section{Conclusion}

The afterglow counterparts of binary neutron star mergers represent viable means to measure the inclination angle of sources, and thereby to improve the standard-siren measurement of the Hubble constant. Afterglows could therefore play the role of narrowing down $H_0$ and possibly resolve the Hubble tension with fewer events than by leveraging only the gravitational-wave data and source redshift. To quantify how much faster afterglow-enhanced $H_0$ measurements could solve the Hubble tension, we carried out a realistic population model considering that every future afterglow counterpart could provide a constraint on the source inclination angle at the same level as GW170817. We found that, while each afterglow allows for a jump in $H_0$ precision, events with afterglow counterparts should prove very rare, to the point that allowing for afterglow-enhanced measurements should not statistically make any difference in the number of events required. Once models have improved, kilonova light curves could be viable for inclination angle measurements and, as these should be much more frequently acquired, kilonovae could play the leading role in multimessenger cosmology. Whether for kilonova or afterglow counterparts, one must treat selection effects with care so as to not produce yet another tension because upcoming gravitational-wave observing runs will probe distances where selections effects are important.

\appendix

\section{Statistical framework for the inferring of $H_0$ with multimessenger data}
\label{sec:appendix}

According to Bayes' theorem under selection effects, the posterior distribution on $H_0$ obtained given a multimessenger set $\textbf{d}$ of GW and electromagnetic data is \citep{2019MNRAS.486.1086M}:
\begin{equation}
    p(H_0|\textbf{d})=\frac{1}{\beta(H_0)}\frac{p(\textbf{d}|H_0)p(H_0)}{p(\textbf{d})}.
    \label{eq:mandel}
\end{equation}
Here, $p(H_0)$ is prior information before the measurement and $p(d|H_0 )$ is the likelihood of the data.
The function $\beta(H_0)$ is generally referred to as selection function or Malmquist bias that accounts for the estimation bias arising from the observation of a distribution in space of astrophysical objects with a flux-limited survey.
The selection function $\beta(H_0)$ corrects for this bias and it reads: 
\begin{equation}
    \beta(H_0)=\int \d \cos \iota \, \d z p_{\rm {\rm det}}(z, \cos \iota | H_0) p_{\rm pop}(z, \cos \iota | H_0),
    \label{eq:selef}
\end{equation}
where $p_{{\rm det}}(z, \cos \iota)$ is the probability of making the measurement on a system located at redshift $z$ and inclination $\iota$ and $p_{pop}(z, \cos \iota | H_0)$ is the distribution of the overall population in redshift and inclination.

As explicit in Eq.~\ref{eq:mandel}, the $H_0$ posterior will only be impacted by selection effects if the selection function is not flat, that is, has a significant variation with $H_0$. Our detection probabilities are best expressed in terms of $d_L$:
\begin{multline}
\beta(H_0) = \int_0^{+\infty} {\rm d} d_L \int_{-1}^{+1} {\rm d} \cos \iota  \frac{\d z}{ \d d_L} p^{\rm L}(d_L, \cos \iota) \\ \times  p_{pop}(z(d_L, H_0), \cos \iota)
\label{eq:explicit}
\end{multline}
with ${\rm L} = {\rm L}_1, {\rm L}_2, {\rm L}_3$ covering our three detection scenario levels.

We consider the formation of binaries uniform in comoving volume \citep{2019PhRvD.100j3523M,2020PhRvD.102d4009M}. Therefore, at low redshift, we have $p_{pop}(z, v) \propto z^2/H_0^3$ 

As explained in Sect.~\ref{sec:3}, due to the small redshift of the sources we consider, we can neglect the effects of redshift on the detected chirp mass and last stable circular orbit for the GW part, and the electromagnetic wavelengths for counterpart searches. Therefore, the detection probabilities $p^{\rm L}$ we consider have no explicit dependence on the source redshift. 
In these circumstances, it is clear from Eq.~\ref{eq:explicit} that $\beta(H_0)$ will be independent of $H_0$ if the cosmology is assumed linear: $\d z = H_0/c\, \d d_L$. Indeed, in this case, the integrand in $\beta(H_0)$ will be:

\begin{multline}
 p^{\rm L}(d_L, \cos \iota) \frac{\d z}{ \d d_L} p_{pop}(z(d_L, H_0), \cos \iota) \\ \propto p^L(d_L, \cos \iota) \frac{H_0}{c} \frac{1}{H_0^3} \left(\frac{H_0 d_L}{c}\right)^2,
\end{multline}
leaving no dependence on $H_0$. 

Therefore, assuming (i) linear cosmology, (ii) uniform-in-comoving-volume system formation rate, and (iii) no explicit redshift dependence in the detection process, the selection effects on the multimessenger measurement of $H_0$ are null. This fact was already underlined in \citet{2019MNRAS.486.1086M}.

\begin{figure*}
    \centering 
    \includegraphics[scale=0.7]{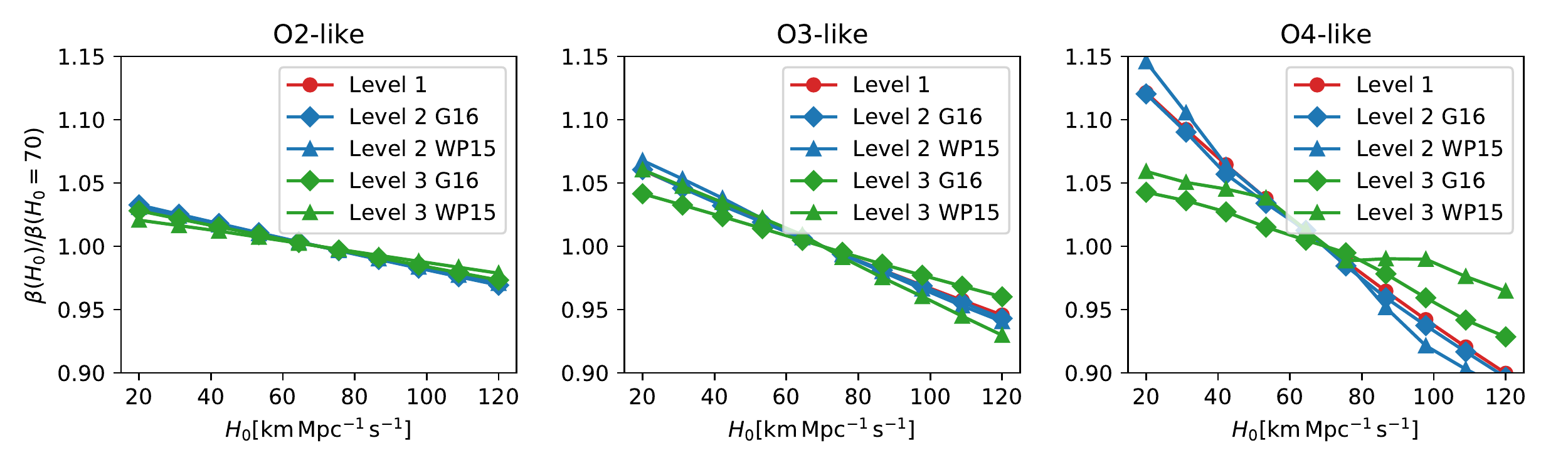}
    \caption{Selection function $\beta(H_0)$ normalized  at an arbitrary nominal value of $H_0 = 40 {\rm km}/{\rm s}/{\rm Mpc}$, for the different observing scenarios we considered.}
    \label{fig:selection}
\end{figure*}

If, however, a general cosmology is considered, such a simplification does not occur and selection effects can appear. In Fig.~\ref{fig:selection}, we plot the selection function for our various observing scenarios assuming a flat Universe with current dark matter density $\Omega_m=0.308$ \citep{PAAAA+2018}. 

\section*{Acknowledgments}
We thank H. Chen for useful comments during the development of this work.
RD acknowledges the Centre National d'Études Spatiales (CNES) for financial support in this research project.
SM is supported by the LabEx UnivEarthS (ANR-10-LABX-0023 and ANR-18-IDEX-0001), of the European Gravitational Observatory, and the Paris Center for Cosmological Physics.
\bibliographystyle{aa}
\bibliography{main}
\end{document}